\documentclass[useAMS,usenatbib,usegraphicx]{mn2e}
\usepackage{color}

\title[A novel variability-based quasar selection method]{A novel variability-based method for quasar selection: evidence for a rest frame $\sim$54 day characteristic timescale\thanks{The data presented herein were obtained at the W.M. Keck Observatory, which is operated as a scientific partnership among the California Institute of Technology, the University of California and the National Aeronautics and Space Administration.}}
\author[M. J. Graham et al.]{Matthew~J.~Graham,$^1$\thanks{E-mail:mjg@caltech.edu} 
S.~G.~Djorgovski,$^1$ Andrew~J.~Drake,$^1$ Ashish~A.~Mahabal,$^1$   
\newauthor
Melissa~Chang,$^1$ Daniel Stern,$^2$ Ciro~Donalek,$^1$
 Eilat~Glikman$^3$\\
$^{1}$California Institute of Technology, 1200 E. California Blvd, Pasadena, CA 91125, USA \\
$^{2}$Jet Propulsion Laboratory, California Institute of Technology, 4800 Oak Grove Drive, Pasadena, CA 91109, USA \\
$^{3}$Department of Physics, Middlebury College, Middlebury, VT 05753, USA
}
\begin{document}

\date{Accepted . Received ; in original form}

\pagerange{\pageref{firstpage}--\pageref{lastpage}} \pubyear{2011}

\maketitle

\label{firstpage}

\begin{abstract}
We compare quasar selection techniques based on their optical variability using data from the Catalina Real-time Transient Survey (CRTS). We introduce a new technique based on Slepian wavelet variance (SWV) that shows comparable or better performance to structure functions and damped random walk models but with fewer assumptions. Combining these methods with {\it WISE} mid-IR colors produces a highly efficient quasar selection technique which we have validated spectroscopically. The SWV technique also identifies characteristic timescales in a time series and we find a characteristic rest frame timescale of $\sim$54 days, confirmed in the light curves of $\sim$18000 quasars from CRTS, SDSS and MACHO data, and anticorrelated with absolute magnitude. This indicates a transition between a damped random walk and $P(f) \propto f^{-\frac{1}{3}}$ behaviours and is the first strong indication that a damped random walk model may be too simplistic to describe optical quasar variability.
\end{abstract}


\section{Introduction}

Quasars have long been known to be variable sources; indeed, \cite{3c48} noted that the most striking feature in optical photometry of 3C 48, one of the first identified quasars, was that the optical radiation varied. Their variability -- photometric, color and spectral -- is well-studied: it is aperiodic in nature, most rapid at the highest energies,  and (anti-)correlated at optical/UV wavelengths with various physical parameters, such as: time lag, rest-frame wavelength, luminosity, radio and emission line properties, the Eddington ratio and estimated black hole mass, e.g., \cite{ulrich, bauer09, mac10, meusinger11, schmidt12, kelly13}. However, the physical mechanisms underlying the (optical/UV) variability remain unclear. It could arise from instabilities in the accretion disk \citep{kawaguchi}, supernovae \citep{aretxaga}, microlensing \citep{hawkins93,hawkins10}, stellar collisions \citep{torricelli}, thermal fluctuations from magnetic field turbulence \citep{kelly09} or more general (Poisson) processes \citep{cidfernandes}.

The utility of quasars in studies of galaxy evolution and black hole growth, e.g., \cite{kollmeier06}, the intergalactic medium, e.g., \cite{hennawi07}, and large-scale structure, e.g., \cite{ross09, clowes13}, have led to a number of large quasar samples, e.g., SDSS \citep{sdssqso} and 2DF \citep{2dfqso}. Historically, these have been constructed based on color selection (with its inherent biases) in the observed visible (rest frame UV) but the new availability of time series data over large areas of sky have led to recent samples based on the unique variability characteristics of quasars: SDSS Stripe 82 \citep{mac11, mac12}, EROS-2 and MACHO \citep{pichara12}, and OGLE II/III \citep{kozlowski10, kozlowski13}. Future samples are expected from PTF and Pan-STARRS (see \cite{palanque11} and \cite{mac12} for predictions), {\it Gaia} \citep{mignard} and, ultimately, LSST \citep{lsst}. 

In these data, the variability of quasars is described either in terms of a structure function \citep{schmidt10} (S10) or as a damped random walk process \citep{kelly09} (K09). Both approaches perform well, achieving reliabilities $>$60\% and completeness levels $>$90\% with SDSS Stripe 82 data compared with SDSS color selection \citep{mac11}. Some authors have also combined them with other variability measures \citep{palanque11, pichara12, kim12}, including color information, to create ensemble predictors. Whilst these may be more efficient, particularly for identifying quasars at intermediate redshifts, i.e. $2.2 < z < 4.0$, e.g., \cite{wu11}, the use of color as a selection feature may reintroduce certain biases that relying solely on variability avoids. We note that these methods have also been used to select blazars based on optical variability \citep{ruan12}.

Variability as a selection technique is free of the potential biases that any spectral (color) based method will have. It thus offers a powerful means of testing the completeness of existing quasar samples, and can lead to  more complete ones. Ironically, variability may be a cause of incompleteness in traditionally-selected samples, especially for highly variable (e.g., beamed) sources: most previous sky surveys were done in an effectively single-pass mode, so if a given source was in a low state at the time, it did not make it into the final catalog. There are now hints that the existing catalogues of blazars may be incomplete by as much as a factor of two due to this effect. This is particularly important for CMB analyses where blazars and flat spectrum radio quasars are the only really significant foreground radio sources and must be fully accounted for.

Aside from correlations with physical parameters, variability is one of the only observational tools available for probing the viscosity of the accretion disk via viscosity dependencies on variability timescales, and constraining the geometry of the corona. Variability may also be the most effect observational discriminator between different accretion states \citep{kelly11}. Such phenomena should manifest as behavioral changes at particular {\it characteristic} scales in the statistics used to describe variability. For example, X-ray variability is well described by a broken power law power spectrum with the scale of the break indicating the size of the X-ray emitting region. Similar characteristic scales are to be expected in the optical as well.

In this work, we present an initial application of variability selection techniques to quasars in the Catalina Real-time Transient Survey (CRTS\footnote{http://crts.caltech.edu}; \cite{drake09, djorgovski12, mahabal11}). This is the largest open (publicly accessible) time domain survey currently operating, covering $\sim33,000$ deg$^{2}$ between $-75^{\circ} < Dec < 70^{\circ}$ (except for within $\sim10^{\circ}-15^{\circ}$ of the Galactic plane) to a depth of $V \sim 19$ to 21.5. Time series exist\footnote{http://www.catalinadata.org} for approximately 500 million objects with an average of $\sim$250 observations over a 8-year baseline. This represents at least a fivefold improvement over SDSS Stripe 82 data in terms of temporal sampling which has so far been the definitive data set for such investigations. Subsequent papers will present an analysis of approximately 200,000 known spectroscopically-confirmed quasars in CRTS and a full sample of variability-selected quasars.

This paper is structured as follows: in section 2, we present the selection techniques that we are considering here, including a new one based on Slepian wavelet variance and in section 3, the data sets we have applied them to. We discuss our results in section 4 and conclusions in section 5. We assume a standard WMAP 7-year cosmology ($\Omega_\Lambda = 0.728$, $\Omega_M = 0.272$, $H_0 = 70.4$ \citep{jarosik}) and our magnitudes are (broadly) on the Vega system.

\section{Variability selection techniques}
\label{techniques}

A number of variability-based features have been employed in the literature for identifying quasars, such as scale and morphological measures, but the two most commonly used in recent analyses are the structure function and modelling the variability as a damped random walk process. We also introduce a new technique for identifying quasars, here, the Slepian wavelet variance.

\subsection{Structure function (SF)}
\label{sfdef}

A structure function is the traditional method for describing quasar variability, either on an individual object basis or in terms of an ensemble e.g., \citet{vandenberk04}. It quantifies the variability amplitude (or rms magnitude difference) as a function of the time lag between compared observations. A number of varying definitions have been used, including: 

\[SF(\Delta t) = \frac{1}{N} \sum_{i=1}^{N}(\Delta m_{ij})^{2} \,\, \mathrm{(Simonetti \,\, et \,\, al. \,\, 1984)} \]

\[ SF(\Delta t) = \left< \sqrt \frac{\pi}{2} | \Delta m_{ij} | - \left< \sigma^2\right> \right> \mathrm{(Vanden \,\, Berk \,\,  et \,\, al. \,\, 2004)} \]

\[SF(\Delta t)  = \mathrm{med} (\Delta m_{ij}^2 ) \,\, \mathrm{(Sumi \,\, et \,\, al. \,\, 2005)} \]

\[ SF(\Delta t) = \left< \sqrt \frac{\pi}{2} | \Delta m_{ij} | - \sqrt {\sigma^2_i + \sigma^2_j} \right> \mathrm{(S10)} \]

\[SF(\Delta t)  = 0.74 (IQR) / \sqrt{N-1} \,\, \mathrm{(Macleod \,\, et \,\, al. \,\, 2012)} \]

\[SF(\Delta t) = \sqrt{\left< \Delta m_{ij}^2 \right> - \left<\sigma^2\right>} \,\, \mathrm{(Bauer \,\, et \,\, al. \,\, 2009)} \]

\[SF(\Delta t) = \sqrt{\left< \frac{\pi}{2} \Delta m_{ij} \right>^2 - \left<\sigma^2\right>} \,\, \mathrm{(Bauer \,\, et \,\, al. \,\, 2009)} \]

\noindent
where $\Delta m_{ij}$ is the magnitude difference between the $i^{th}$ and $j^{th}$ observations $(i < j)$
whose time lag ($t_j - t_i$) falls in the bin $\Delta t$ and with measurement uncertainties of $\sigma_i$ and $\sigma_j$ respectively. $<\ldots>$ denotes an ensemble average, med the median and $IQR$ is the interquartile range (the difference between the 75$^{th}$ percentile and the 25$^{th}$ percentile).

The SF is then characterized via a simple power law parameterization (S10)):

\begin{equation}
\label{sfeqn}
SF = A \left( \frac{\Delta t_{obs}}{1 \,\, \mathrm{year}} \right)^{\gamma} 
\end{equation}

\noindent 
where $A$ quantifies the rms magnitude difference on a one year timescale and $\gamma$ is the logarithmic gradient of this mean change in magnitude. This can be fit to the data by maximizing the associated (log) likelihood function, either by Markov Chain Monte Carlo (MCMC) or a standard optimization algorithm \citep{palanque11}. A small $\gamma$ indicates periodic or white noise variability whilst a large value is associated with secular or random-walk-like variability.

Using Stripe 82 data, S10 defined a region in the $A-\gamma$ plane to select quasars, although they constrained its MCMC fits such that $0 < A < 1$ and $\gamma > 0$. \cite{palanque11} trained an artificial neural network on these $A$ and $\gamma$ to estimate quasar likelihood. We have implemented a support vector machine (SVM) on $A$ and $\gamma$ to find the separating hyperplane between the quasars and non-quasars in this parameter space. 

\subsection{Damped random walk (CAR(1))}

Quasar light curves are seen to be aperiodic and noisy with power spectra that lack peaks: this suggests that they are stochastic or chaotic in nature. K09 proposed that they could therefore be modelled as a damped random walk or Ornstein-Uhlenbeck (OU) process, a particular type of (Gaussian) first-order continuous time autoregressive process (CAR(1)) (although we note that \cite{vio92} first suggested describing quasar variability via stochastic differential equations). The temporal behaviour of the quasar flux $X(t)$ is formally given by:

\[ dX(t) = - \frac{1}{\tau} X(t) dt + \sigma \sqrt{dt} \epsilon(t) + b dt, \,\,\, \tau, \sigma, t > 0 \]

\noindent
where $\tau$ is the relaxation time of the process or the time for the time series to become roughly uncorrelated, $\sigma$ is the variability of the time series on timescales short compared to $\tau$, $b\tau$ is the mean magnitude, and $\epsilon(t)$ is a white noise process with zero mean and variance equal to 1. The corresponding likelihood function involves an exponential covariance matrix:

\begin{equation}
\label{car1eqn}
S_{ij} = \frac{\tau \sigma^{2}}{2} \exp(- | t_i - t_j | / \tau) 
\end{equation}

\noindent
which has a tridiagonal inverse and so all calculations are linear in scope (${\cal O} (N_{\mathrm{data}})$) \citep{rybicki95}. Again both MCMC (K09) and optimization \citep{pichara12} techniques have been applied to derive the best-fit parameters. 

Using Stripe 82 data supplemented with additional photometry from SDSS and POSS, \cite{mac12} have shown that a CAR(1) model is a viable description of the optical continuum variability of quasars on timescales of $\sim5 - 2000$ days in the rest frame. However, {\it Kepler} and OGLE data suggests that, on shorter timescales, correlations become stronger than predicted by CAR(1) \citep{mushotzky11, zu13}. \cite{kelly11} also find that 13\% of optical light curves in a study of 55 MACHO AGN are better described by a mixed OU process with multiple characteristic timescales rather than a single one. Finally, \cite{andrae13} considered a variety of different stochastic models with 6304 Stripe 82 quasars, including several extensions to the OU process, but find that the (Gaussian) CAR(1) model is still the best descriptor. There may, however, be a bias toward other models for QSO light curves with fewer observations and at lower redshifts.

Quasars have been selected by defining appropriate regions in the $\tau - \sigma$ plane \citep{kozlowski10, mac11}. \cite{butler11} (BB11) derived linear relationships between $\log (\tau)$ and $\log(\sigma)$ and the median magnitude of a light curve from the ensemble structure functions of quasars in a data set, in this case Stripe 82, predicted by a CAR(1) model ($SF \propto \sigma \tau^{1/2} [ 1 - \exp (-\tau_{ij} / \tau)]^{1/2}$) and then determined how well a CAR(1) process with these best-fit values described a given light curve. Meanwhile, \cite{pichara12} used CAR(1) features plus other time series features to feed a boosted random forest classifier. We consider both selection regions within the plane and the BB11 approach with appropriate relationships for our data sets.

\subsection{Slepian wavelet variance (SWV)}

Wavelets are a popular tool in the analysis of time series. In the same way that Fourier terms can be used to decompose the variance of a time series process with respect to frequency (via the power spectrum), they can be used to decompose the sample variance of a time series on a scale-by-scale basis, e.g., \citet{percival, scargle93}. This means that the relative contributions of variations operating over a range of different time scales to particular physical phenomena can be identified, facilitating studies of period changes, quasi- and multiperiodicity and characteristic time scales amongst others. Standard wavelet analyses (estimating the variance as a function of time scale) assume a regularly sampled time series and so specific approaches have had to be developed to deal with the irregularity common to astronomical time series. Most astronomical applications of wavelets have also dealt with stellar data, although \citep{dong10} have analyzed long-term variability data of 3C 345 using the weighted wavelet Z-transform technique \citep{foster96}. 

Although there are a range of wavelet types that can be used, e.g., Haar, Debauchies, etc., \cite{mondal11} have shown how Slepian wavelets can be applied to irregular and gappy time series (see Appendix~\ref{appa} for details) which makes them ideal for studying astronomical light curves. These can be regarded as optimal approximations to ideal bandpass filters used in estimating wavelet variance. Generally, an observed time series $X_t$ can be decomposed by applying a set of wavelet filters $\{h_{jl}\}$:

\[ W_{j,t} = \sum_{l=0}^{L_j -1} h_{jl} X_{t-l}; \,\, t = 0, \pm 1, \ldots; \,\, j = 1, 2, \ldots; \,\, L \ge 2d \]

\noindent
with the wavelet coefficient, $W_{j,t}$ proportional to changes in adjacent weighted averages over an effective scale of $2^{j-1}$ time points. The wavelet variance at scale $\tau_j = 2^{j-1}\bar{\Delta}$ where $\bar{\Delta}$ is the average sampling interval is then:

\[ \nu_X^2(\tau_j) = \mathrm{var}(W_{j,t}) \]

\noindent
and gives the contribution to the total variance of the time series due to scale $\tau_j$ ($\mathrm{var} (X_t) = \sum_{j=1}^{\infty} \nu_X^2(\tau_j)$)\footnote{\cite{scargle93} uses the value of the mean squared wavelet coefficient rather than its variance.}  This means that large values of $\nu_X^2$ compared to surrounding values indicate characteristic scales (either local or global, depending on the magnitude of the variance) \citep{keim10}. 

In fact, given the power spectrum $P(f)$ of a time series, the wavelet variance is approximately given by:

\[ \nu_X^2(\tau_j) \simeq \int P(f) \,\, \mathrm{d}f \]

\noindent
where the integral is over the dyadic frequency band associated with $\tau_j$ (the exact expression also involves the squared gain function of the wavelet filter but this is not necessarily analytic). Thus features in the SWV distribution correlate with particular patterns of behavior at particular scales. A periodic scale, $P$, is expected to manifest as a peak in the variance near the scale $P/2$. Alternatively, a linear relationship between $\log (\nu_X^2)$ and $\log (\tau)$ over a range of $\tau$ might suggest power law behaviour over those scales ($P(f) = f^\alpha \Rightarrow \nu_X^2(\tau) \propto \tau^{-(\alpha + 1)}$). Finally, a CAR(1) process has a Lorentzian power spectrum: 

\[ P(f) = \frac{\sigma^{2}}{(2 \pi)^3} \frac{1}{f_0^2 + f^2} \]

\noindent
where $\tau = 1 / (2 \pi f)$ using the same notation as eqn.~(\ref{car1eqn}). This gives a wavelet variance of:

\[ \nu_X^2(\tau_j) \propto  \frac{\sigma^2 \tau_0}{2 \pi}  \arctan \left( \frac{\tau_0}{\tau} \right) \]

\noindent
Note that a light curve needs a minimum of 12 observations to determine its SWV values.

Changes between relationships, e.g., different slopes, indicate transition regions between behaviors -- characteristic timescales corresponding to break frequencies in the power spectrum, say. Obviously, this allows for a far more direct investigation of the scale dependencies of the various processes contributing to quasar variability and characteristic scales indicating changes of behavior than other techniques where they have to be inferred. However, unlike structure function and damped random walks, no actual assumptions are made about the forms of any underlying processes. This could mean potential contamination by other classes of object with similar behaviours and/or transitions (characteristic timescales) to quasars, e.g., long period variables (LPVs), but the same can be argued for both SF and CAR(1).

\section{Data sets}
There are now a number of data sets with sufficient sky and/or temporal coverage to produce large $(n > 1000)$ catalogs of quasar candidates. CRTS will produce the largest known quasar variability sample with a conservative estimate of over a million quasar candidates (comparable to the total number of SDSS photometric quasar candidates). As a check on our selection criteria, we have also used the SDSS Stripe 82 data set.

\subsection{Catalina Real-time Transient Survey (CRTS)}
\label{crts}

CRTS leverages the Catalina Sky Survey data streams from three telescopes used in a search for Near-Earth Objects, operated by Lunar and Planetary Laboratory at University of Arizona. CRTS covers up to $\sim$2500 deg$^2$ per night, with 4 exposures per visit, separated by 10 min., over 21 nights per lunation. All data are automatically processed in real-time, and optical transients are immediately distributed using a variety of electronic mechanisms\footnote{http://www.skyalert.org}. 

The full CRTS data set contains time series for approximately 500 million sources. We have undertaken an initial study of 1.5 million sources from the first CRTS data release\footnote{http://nesssi.cacr.caltech.edu/DataRelease/index1.html} of 200 million sources to determine how to most efficiently characterize such time series (Graham et al., in prep.). This data set consists of a sample of 1 million variable sources and a control sample of 500,000 sources which have been deemed to be non-variable. For each sky region (frame) covered by CRTS, the most variable sources have been determined from the Stetson $J$ values \citep{stetson96} of their time series as a function of magnitude and number of observations. An exponential weighting scheme between successive pairs of observations in time order was used. The list of sources was then randomly sampled to select a million sources. A control sample of a further 500,000 sources was also randomly selected from those objects which were not selected as (highly) variable.

\begin{figure}
\caption{The $V$-band magnitude distribution for the CRTS known quasar,  control and stellar samples (red histogram). The solid line shows the SDSS $r$-band magnitude distribution for the S82 sample.}
\label{mag}
\includegraphics[width=3.3in]{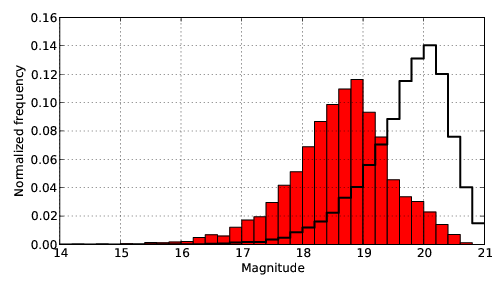}
\end{figure}

\begin{figure}
\caption{The distribution of the number of observations in a time series for the CRTS known quasar, control and stellar samples (red histogram). The solid line shows the same distribution for the S82 sample. }
\label{obs}
\includegraphics[width=3.3in]{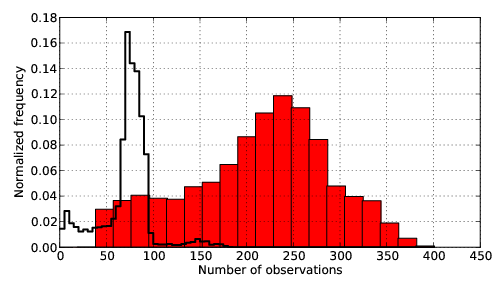}
\end{figure}

\begin{figure*}
\caption{The distribution of the two CRTS data sets on the sky in Galactic coordinates: (left) the 7300 known quasars and (right) the control sample of 7113 non-quasars. The southern sky artifacts in the quasar distribution indicate the SDSS origin of most of the objects. The black region in both plots is the distribution of S82 data.}
\label{sky}
\includegraphics[width=7.0in]{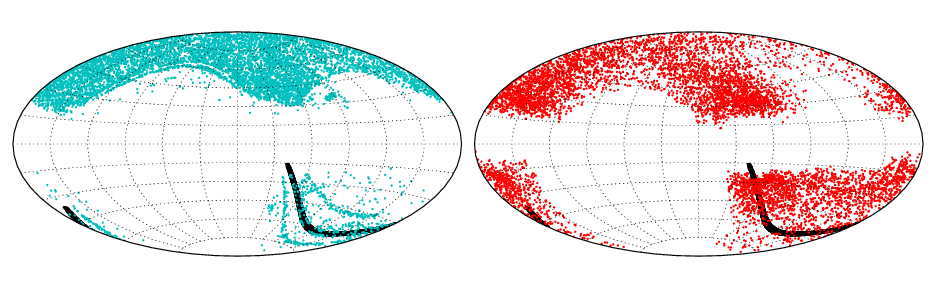}
\end{figure*}

\begin{figure}
\caption{The redshift distribution for the CRTS known quasar sample (red histogram), omitting two quasars at 4.73 and 5.657. The solid line shows the same distribution for the S82Q sample.}
\label{redshift}
\includegraphics[width=3.3in]{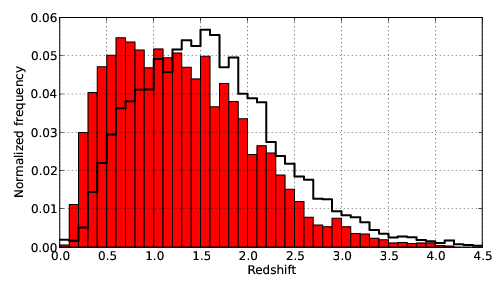}
\end{figure}

The Million Quasars (MQ) catalogue\footnote{http://quasars.org/milliquas.htm} v3.5 contains all spectroscopically-confirmed type 1 QSOs (306,687), AGN (22,552) and BL Lacs (1750) in the literature up to 10 August 2013. We have crossmatched this against our data sets with a 3\arcsec matching radius and find that about 191,000 confirmed quasars (``Q'' designations in MQ) are covered by the full CRTS. A subset of 7300 quasars are found in our initial study data set and forms the basis for this analysis. We have also selected a control sample of 7113 sources which are not known to be quasars or quasar candidates, i.e., they have no match in MQ, with the same magnitude distribution and coverage (no. of observations) as the quasar sample (see Figs.~\ref{mag}, \ref{obs} and \ref{sky}).  The redshift distribution for the quasars is shown in Fig.~\ref{redshift}.

In a variability-based search for QSOs behind the LMC, \citep{geha03} found that Be/Ae stars were the main source of contamination. Above the Galactic plane, \citep{kozlowski10} found that Cepheids, RR Lyrae, some LPVs, eclipsing binaries, slowly pulsating Bs, $\beta$ Cepheids and active giant and subgiants all overlapped the CAR(1) quasar locus area defined with OGLE data. In analyzing probable contaminants, S10 considered F/G stars and RR Lyrae for Stripe 82 data. We have constructed a sample of 4356 stars (3404 RR Lyrae, 952 other classes) with the same normalized joint magnitude and number of observations distribution as the quasar sample from the set of RR Lyrae and variable stars reported in \cite{drake13} and \cite{graham13}, respectively.

\subsection{Stripe 82 (S82)}

SDSS Stripe 82 (e.g., \cite{sesar07}) is one of the most repeatedly observed large ($>100 \deg^2$) regions of the sky and thus has served as the basis for many studies of variable behaviour in astronomical populations (e.g., \cite{suveges12}). A sky region defined by $22^h24^m < RA < 04^h08^m$ and $-1.27^\circ < Dec < 1.27^\circ$ (covering an area of $\sim$290 deg$^2$) was initially imaged about once to three times a year from 2000 to 2005 (SDSS-I) and then with an increased cadence of 10--20 times a year from 2005 to 2008 (SDSS-II SNe search), giving an average sampling of 53 epochs with a baseline of 5 to 8 years in SDSS {\it ugriz}-bands.

We have retrieved the SDSS Stripe 82 data for 11150 sources (hereafter referred to as S82), consisting of 10293 objects spectroscopically classed as QSOs in SDSS DR10 \citep{ahn13} (hereafter referred to as S82Q) and 957 objects with either another spectroscopic class (288) or none at all (669). For the purposes of this analysis, we will only consider the SDSS $r$-band time series data for each object. We note that there are 62 quasars in common between the CRTS known quasar sample and S82Q.

\subsection{Preprocessing}
It is common to preprocess data to remove spurious outlier points which may be caused by technical or photometric error. The danger, of course, is removing real signal, although a robust method would be unaffected by the presence of noisy data. We have created a cleaned version of each data set following the procedure of \citep{palanque11} (PD11): a 3-point median filter was first applied to each time series, followed by a clipping of all points that still deviated significantly from a quintic polynomial fit to the data. To ensure that not too many points are removed, the clipping threshold was initially set to 0.25 mag and then iteratively increased (if necessary) until no more than 10\% of the points were rejected. Note that PD11 used a 5$\sigma$ threshold but we found that this removed too few points and so used the limit set by S10 instead.

\section{Results}
We have applied the three techniques described in Sec.~\ref{techniques} to the CRTS  and S82 quasar and stellar data sets. We defer discussion of the CRTS control sample to the next section.

To quantify the performance of the various quasar selection methods, we use the completeness $C$ and purity $P$:

\[
C = \frac{\mathrm{Number \,\, of \,\, selected \,\, quasars}}{\mathrm{Total \,\, number \,\, of \,\, confirmed \,\, quasars}}\]

\[
P = \frac{\mathrm{Number \,\, of \,\, selected \,\, quasars}}{\mathrm{Total \,\, number \,\, of \,\, selected \,\, objects}}\]

\noindent
as well as the F-score, $F_1 = 2 (C \times P) / (C + P)$, which provides a single value statistic to aid comparison.

\subsection{Structure function}
We have fitted the parameterized form of the SF (eqn.~\ref{sfeqn}) to both data sets according to S10 using both a MCMC approach (employing the Python {\bf emcee} module \citep{foreman} which provides an affine-invariant ensemble sampler) and optimization using a non-linear Nelder-Mead algorithm (which gave better convergence rates than the popular Minuit algorithm). For the MCMC, we assumed the same uninformative priors for the fitted parameters as S10 (uniform in the logarithm of A and the arctangent of $\gamma$):

\[
p(A) \propto \frac{1}{A}, \qquad p(\gamma) \propto \frac{1}{1 + \gamma^2}.
\]

\begin{figure*}
\caption{The distribution of $A$ and $\gamma$ parameters derived from optimization fits of a parameterized structure function to the CRTS (left) and S82 (right) data sets. The cyan points indicate quasars in both data sets and red points the control sample (stars and non-quasars) respectively. The S82 distribution is consistent with Fig.~6 in P11. The black line indicates the quasar selection region defined in S10.}
\label{sffig}
\includegraphics[width=7.0in]{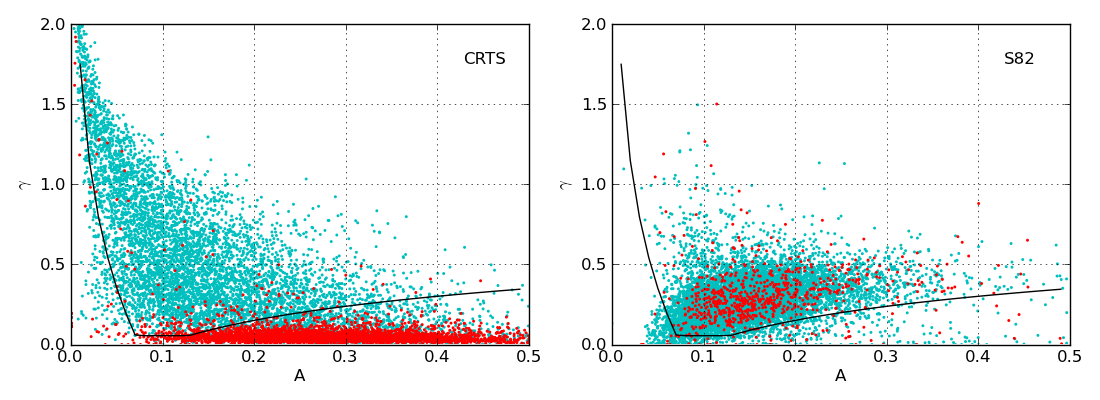}
\end{figure*}

\noindent
$A$ and $\gamma$ are also constrained to: $A \in [0,1], \gamma > 0$. The results from both approaches were equivalent but the optimization process was found to be faster. The respective distributions for the data sets are shown in Fig.~\ref{sffig}. S10 define a quasar selection region in the A-$\gamma$ plane. The respective confusion matrices are given in Table~\ref{rrcon}. We note that we also tried the allowed parameter ranges used by P11 ($A \in [0, 5], \gamma \in [-0.1, 10]$) but found that these gave poorer results.

Using the S10 quasar selection region, we find a completeness of 74\% and a purity of 96\% for CRTS data and a completeness of 92\% and a purity of 93\% for the S82 data. The lower results for CRTS data are probably due to a combination of a number of factors. There are differences in time sampling between the two data sets: S10 find a completeness of 93\% for their S82 data but note that with different (Pan-STARRs 1-like) time sampling, this would drop to 76\%. CRTS data is also noisier than S82 data and although the cleaning procedure used with both should mitigate these effects, we find that the $(A, \gamma)$ results can easily vary by 10-20\%, depending on how many points are rejected. 

For comparison, we have trained a support vector machine (SVM) on the CRTS data and applied this to S82. Selecting quasars via the trained SVM gives completeness and purity values of 77\% and 80\% for CRTS and 77\% and 92\% for S82. Again, the differences in time sampling and noise levels between the two data sets are likely contributing factors.

\begin{table*}
\caption{The confusion matrices derived from the various algorithms to identify quasars and stars in CRTS and quasars and non-quasars in Stripe 82 data. Note that two S82 light curves had fewer than 12 points and so do not have a SWV value. Note also that completeness, purity and $F_1$-scores for BB11 do not include objects with ambiguous classifications.}
\label{rrcon}
\centering
\begin{tabular}{lllllllll}
\hline
CRTS & \multicolumn{2}{c}{\bf{SF}}  & \multicolumn{2}{c}{\bf{CAR(1)}} & \multicolumn{2}{c}{\bf{BB11}} & \multicolumn{2}{c}{\bf{SWV}} \\
 & QSO & Star  & QSO & Star & QSO & Star & QSO & Star \\
\hline
QSO & 5157 & 2143  & 6651 & 649 & 4915 & 2385 & 6775 & 525\\
Star & 200 & 4155 & 3442 & 914 & 1950 & 2406 & 1664 & 2692 \\
\hline
Completeness & \multicolumn{2}{c}{71\%} & \multicolumn{2}{c}{91\%} & \multicolumn{2}{c}{67\%} & \multicolumn{2}{c}{93\%} \\
Purity & \multicolumn{2}{c}{96\%} & \multicolumn{2}{c}{66\%} & \multicolumn{2}{c}{72\%} & \multicolumn{2}{c}{80\%}\\
$F_1$-score & \multicolumn{2}{c}{0.81} & \multicolumn{2}{c}{0.76} & \multicolumn{2}{c}{0.69} & \multicolumn{2}{c}{0.86}\\
\hline
\hline
 S82 & QSO & Non-QSO  & QSO & Non-QSO & QSO & Non-QSO & QSO & Non-QSO\\
\hline
QSO & 9486 & 807  & 9303 & 390 & 9687& 606 & 8852 & 1399\\
Non-QSO & 765 & 92 & 813 & 44 & 815 & 42 & 711 & 86 \\
\hline
Completeness & \multicolumn{2}{c}{92\%} & \multicolumn{2}{c}{96\%} & \multicolumn{2}{c}{94\%} & \multicolumn{2}{c}{86\%} \\
Purity & \multicolumn{2}{c}{93\%} & \multicolumn{2}{c}{92\%} & \multicolumn{2}{c}{92\%} & \multicolumn{2}{c}{92\%}\\
$F_1$-score & \multicolumn{2}{c}{0.92} & \multicolumn{2}{c}{0.94} & \multicolumn{2}{c}{0.93} & \multicolumn{2}{c}{0.89}\\
\hline
\end{tabular}
\end{table*}

\subsection{CAR(1)}
We have fitted CAR(1) processes to both data sets again using both a MCMC and an optimization approach and exploiting the tridiagonal inverse formalism in each case. The parameters of interest are $\tau$ and $\sigma^2$ ($b$ is estimated as part of the fitting). We note that alternate techniques have been used to fit CAR(1) models using Kalman iterative relationships (e.g., K09) or Bayesian techniques (e.g, \citep{andrae13}) and a direct approach applicable to general CAR(p) processes is also discussed in \citep{koen05}. For the MCMC, we have assumed ``data-based'' priors (\citep{bailer-jones13}) with a gamma prior on both $\tau$ and $\sigma^2$ with shape = 1.5. The scale of the $\tau$ prior is set to one quarter of the duration of the time series in question and that of $\sigma^2$ to $2 \sigma^2_x / \tau$ where $\sigma_x^2$ is the variance of the time series. 

\begin{figure*}
\caption{The distribution of $\tau$ and $\sigma^2$ parameters derived from MCMC fits of a CAR(1) process to the CRTS (left) and S82 (right) data sets. The cyan points indicate quasars in both data sets and red points indicate the control (non-quasars) sample. The S82 distribution is consistent with Fig.~3 in BB11. The black line indicates the quasar selection region defined in K09.}
\label{carfig}
\includegraphics[width=7.0in]{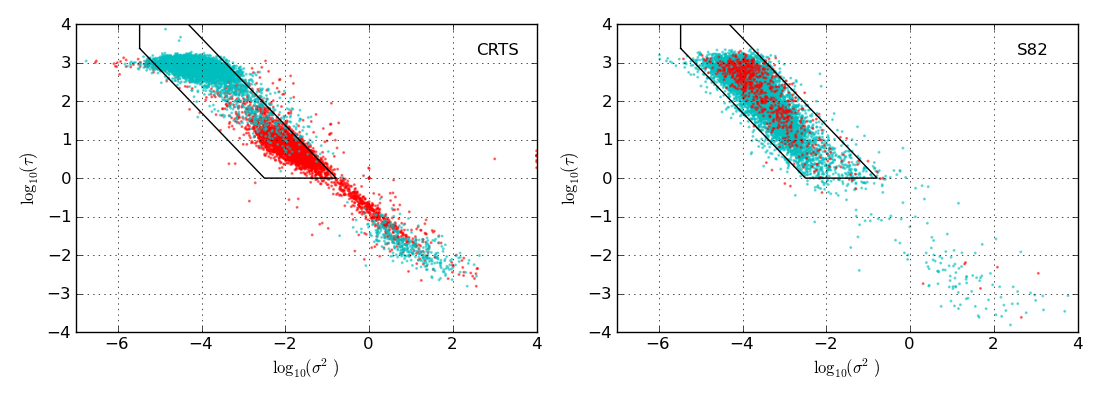}
\end{figure*}

The respective distributions of $\tau$ vs. $\sigma^2$ for both data sets are shown in Fig.~\ref{carfig}. We have also determined the appropriate coefficients for the BB11 method for CRTS data from its ensemble structure function. We find: $a_1 = -3.37, a_2 = 0.42, a_3 = 2.82$ and $a_4 = -0.02$ where $\log(\sigma^2) = a_1 + a_2(\mathrm{mag} - 19)$ and $\log(\tau) = a_3 + a_4(\mathrm{mag} - 19)$. The corresponding values for SDSS $r$-band data are: $a_1 = -4.34, a_2=0.20, a_3 = 3.12$ and $a_4=-0.15$.

Using the type of quasar selection region defined in K09, we find a completeness of 91\% and a purity of 66\% for CRTS data (see Table~\ref{rrcon}). The respective quantities for S82 data are 96\% and 92\%. The completeness for the CRTS data is better than with the SF but the purity is much reduced. As Fig.~\ref{carfig} shows, the purity could be improved -- $\log(\tau) > 1.0$ rather than $\log(\tau) > 0.0$, say, giving a CRTS purity of $85\%$ with a slightly lower completeness but this would affect the S82 results with a reduced completeness of $88\%$ and purity of 93\%. It is clear, therefore, that the vast majority of the S82 stars lie within the quasar selection region and so with a more balanced set, the purity may well be substantially lower.

The BB11 method produces poorer results with a completeness of 67\% and a purity of 72\% for CRTS data; however, these statistics do not include the 911 quasars and 26 stars that were classified as ``ambiguous''.
It seems likely that the ensemble structure function used in BB11 is better constrained for S82 data than for CRTS data and that this then leads to more uncertainty or actual misclassification in the fit. We find that we can obtain better numbers for the CRTS data (fewer ambiguous classifications) if we artificially reduce the scale of the photometric errors or equivalently get worse numbers for S82 if we increase the error size. Although an advantage of BB11 over the other techniques is its speed -- just one goodness-of-fit calculation rather an optimization, for noisier data, such as CRTS, individualized fits give better results than a global approach.

\subsection{Slepian wavelet variance}
\label{swv}

\begin{figure*}
\caption{The distribution of $\tau$ and Slepian wavelet variance for all objects in the CRTS quasar (left) and stellar (right) samples. The black points indicate the median variance value used to determine class membership with error bars corresponding to the median absolute deviation from median.}
\label{swvcls}
\includegraphics[width=7.0in]{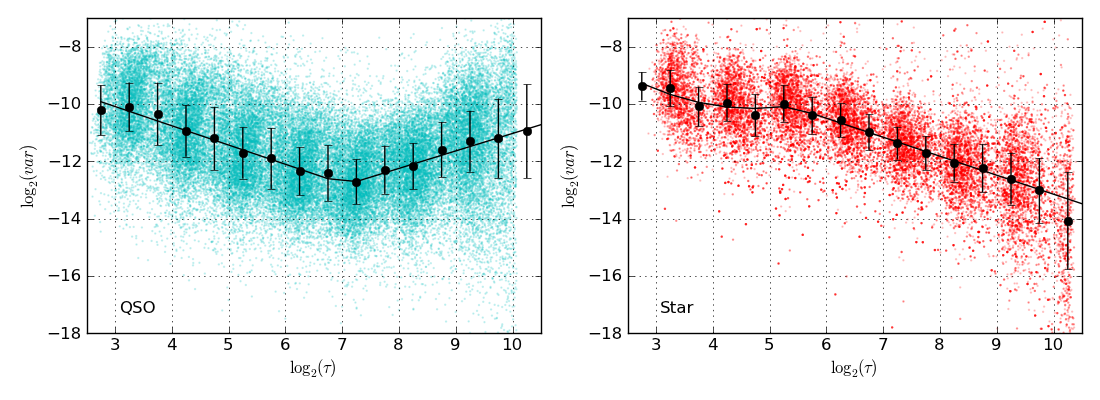}
\end{figure*}

We have determined the Slepian wavelet variance (SWV) distribution for all data sets.  Fig.~\ref{swvcls} shows the respective distributions for the CRTS quasars and stars. There is a clear difference in shape between the two and we can use this to identify quasars. We have median binned each distribution and calculated optimal least squares fits to these; we find that quasars follow:

\[
\log_2(var) = \left\{ \begin{array}{rl}
 -0.672 x - 8.073, & x < 7.064 \\
0.610 x - 17.130, & x > 7.064 \end{array} \right.
\]

\noindent
and stars follow:

\[
\log_2(var) = \left\{ \begin{array}{ll}
 0.225 x^2 -2.127 x - 5.134, & x \le 5.25 \\
-0.663 x - 6.520, & x > 5.25 \end{array} \right.
\]

\noindent
where $x = \log_2(\tau)$. To determine whether an object is a quasar or not, we fit both expressions to the SWV curve of the object, allowing an additive scaling factor in $\log-\log$ space to both components in each expression, i.e., we fit $-0.672x -8.073 + a$ and $0.610x -17.130 + b$ for quasars and similarly for stars. This reflects the fact that the broad shape of the distribution determines the class but it needs to be appropriately scaled to the level of each object (this reflects a magnitude dependency (see Sec.~\ref{timescales}). The fit with the better reduced chi-squared determines the class, i.e., $\chi^2_{qso} < \chi_{star}^2$ is a quasar and vice versa.  

\begin{figure}
\caption{The distribution of $\tau$ and Slepian wavelet variance for all Stripe 82 quasars. The black line denotes the median fit to the CRTS quasar distribution scaled as described in the text.}
\label{swv82}
\includegraphics[width=3.3in]{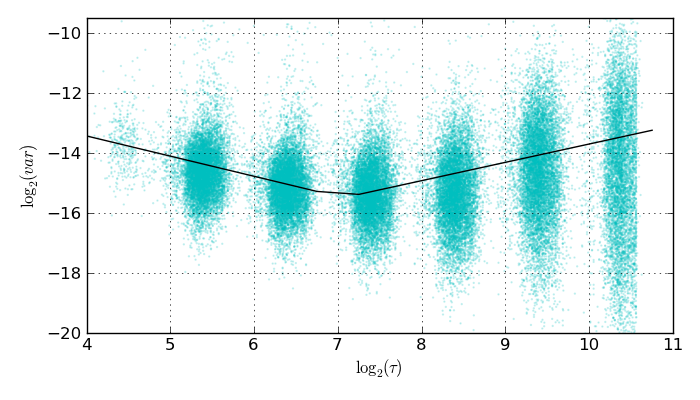}
\end{figure}

We find a completeness of 93\% and a purity of 80\% for the CRTS data and a completeness of 86\% and a purity of 92\% for the S82 data (see Fig.~\ref{swv82}). The distinction between a quasar and a star is more apparent in the CRTS data than in S82 and this is almost certainly due to the better time coverage of CRTS data. The SWV is calculated on a dyadic timescale (see Appendix~\ref{appa} for details) with $\tau_j = \bar{\Delta} 2^{j-1}, j = 1, \ldots, \log_2(N) - 1$ so for S82 data with a median number of 75 observations and a median average sampling interval of $\bar{\Delta} = 45.2$ days, SWV will typically consist of 5 points over a range $5.5 \le \log_2(\tau) \le 9.5$. For CRTS, however, $\bar{\Delta} = 9.9$ days with a median number of 221 observations and so SWV will typically consist of 7 points over a range 
$3.3 \le \log_2(\tau) \le 9.3$. SWV therefore has greater discriminative power with CRTS data over S82 data.

\subsection{Dependencies}

\begin{figure*}
\caption{The completeness and purity of each algorithm as a function of magnitude and number of observations: structure function (left), CAR(1) (middle) and Slepian wavelet variance (right). The black lines in the histograms indicate the normalized cumulative distribution of the respective variable. The minimum number of observations in a CRTS light curve in this data set is 40. There are also no brighter quasars than $V \sim 15$.} 
\label{compdep}
\includegraphics[width=2.15in]{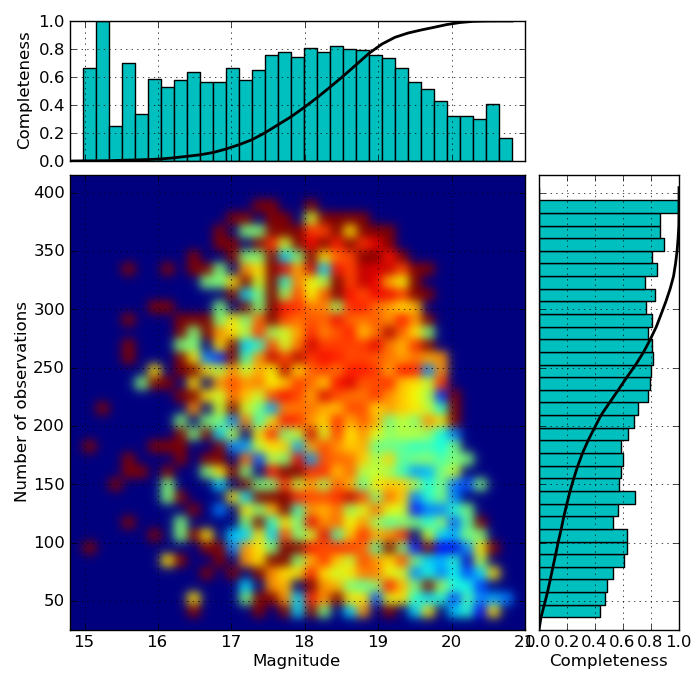}
\includegraphics[width=2.15in]{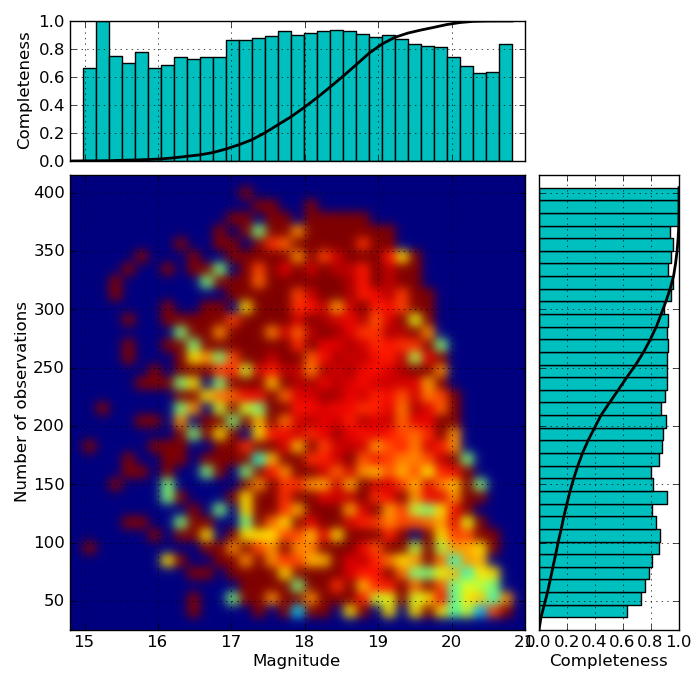}
\includegraphics[width=2.15in]{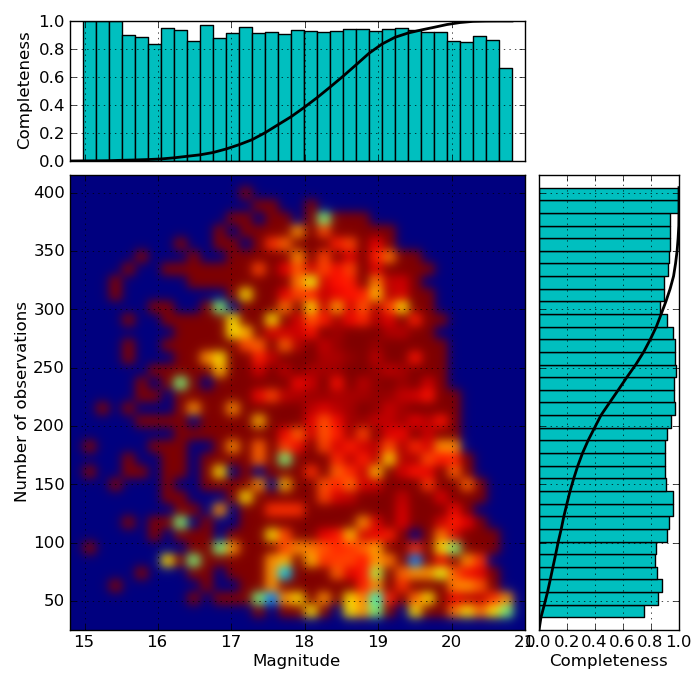}
\includegraphics[width=0.32in]{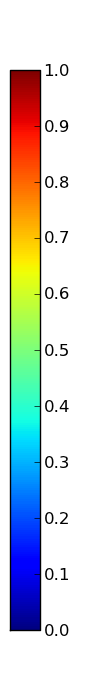} \\
\includegraphics[width=2.15in]{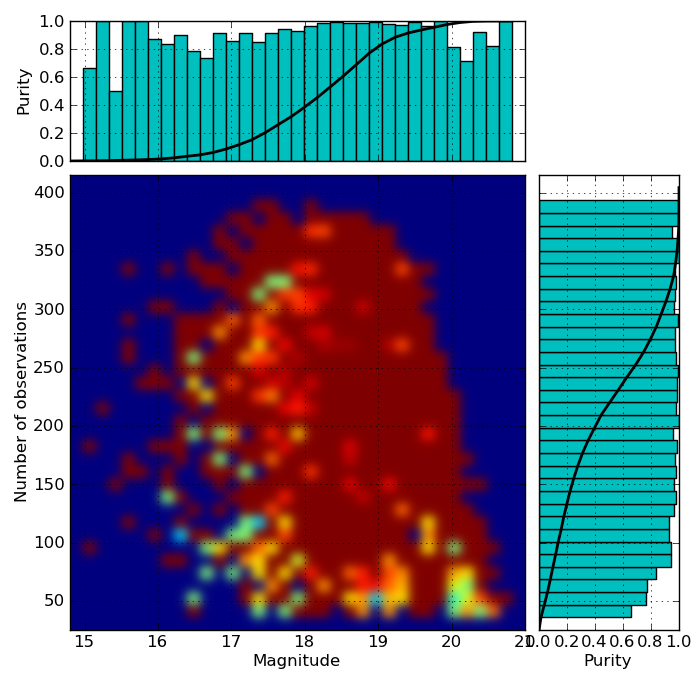}
\includegraphics[width=2.15in]{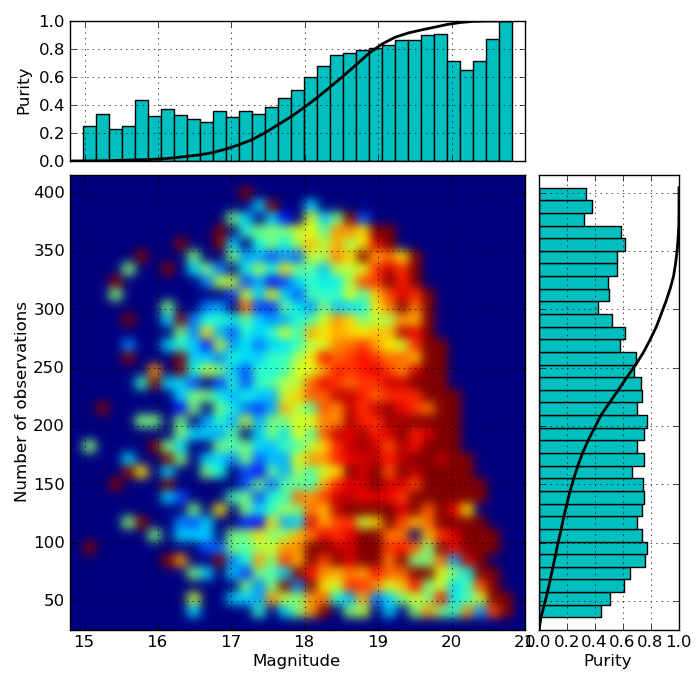}
\includegraphics[width=2.15in]{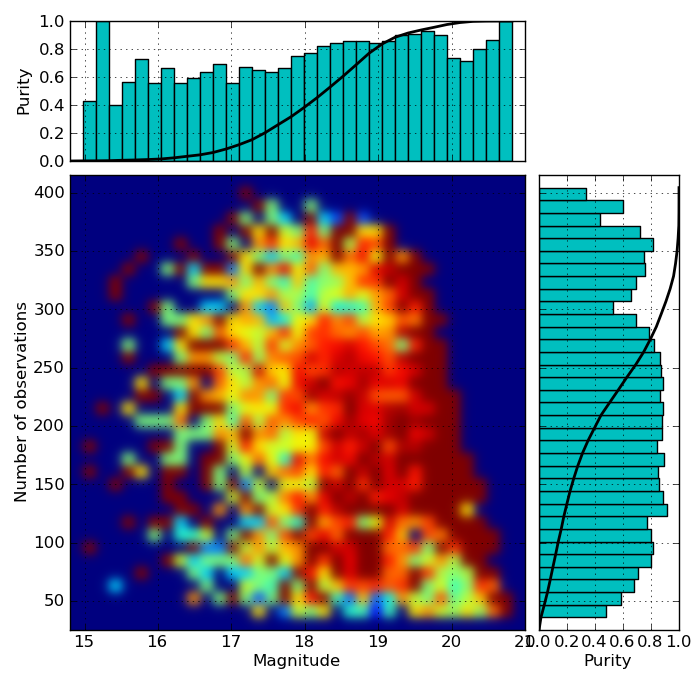}
\includegraphics[width=0.32in]{control_cb}
\end{figure*}

It is clear that the different quasar selection algorithms considered here have varying efficiencies depending on the quality and nature of the data. Although we defer a systematic analysis relating the variability dependencies of quasars to physical parameters to a subsequent paper, we can at least examine here how these broadly correlate with the quality of the data. The relative completeness and purity of each algorithm as a function of magnitude (signal) and number of observations (sampling) is shown in Fig.~\ref{compdep}. 

\begin{figure}
\caption{The amplitude of quasar variability (as measured by the median absolute deviation from the median) as a function of magnitude for the CRTS quasars.}
\label{varmag}
\includegraphics[width=3.3in]{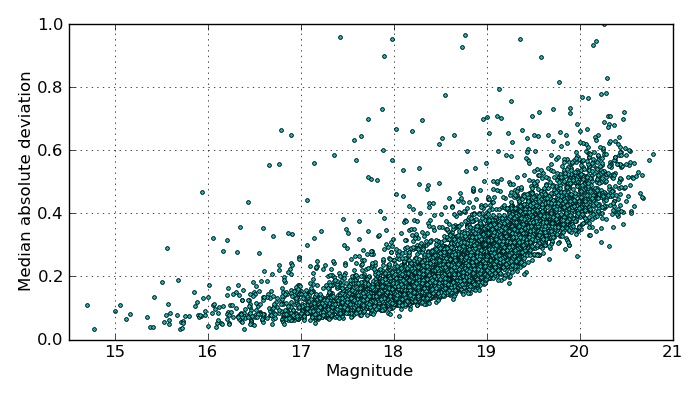}
\end{figure}

For all three algorithms, there is a slight trend of increased completeness with greater number of observations (most notably shown by SF) but purity seems rather flat in comparison. This obviously relates to better sampling or a finer resolution of quasar behaviour as the number of observations increases. As far as magnitude is concerned, SF completeness and CAR(1) and SWV purity all show strong dependencies. The amplitude of quasar variability is strongly correlated with magnitude (see Fig.~\ref{varmag}) and so at fainter magnitudes, the quasar signal is more easily discernible (in spite of larger photometric errors): there are fewer misclassifications and the purity increases. However, SF has a much stronger dependency on photometric error (see the statistic definitions in Sec.~\ref{sfdef}) which dominates at the faintest magnitudes, despite the strong variability signal.

\cite{mac10} note that scatter in the fitting parameters for CAR(1) can also be introduced by insufficient time sampling, although the argument holds for all the algorithms considered here. Approximately 60\% - 70\% of the spread in the CAR(1) parameters ($\tau$ and $\sigma^2$) is attributable to sparse sampling and limited baselines in S82 light curves. As shown in Fig.~\ref{obs} and discussed in Sec.~\ref{swv}, CRTS has better sampling by a factor of $\sim$5 than S82 but over a shorter timescale (the median timescale for CRTS is 2208 days whereas for S82 it is 3343 days). This should mean that there is less scatter in the fitted parameters for CRTS data than for S82 and this does generally seem to be the case with all the algorithms. Another assumption is that light curves are long enough to get accurate estimates for timescales and asymptotic parameters. Neither of the SF variables ($A$ and $\gamma$) are of this type and so it is unaffected by this argument. \cite{mac12} identify a characteristic timescale of $\sim$1400 days from a CAR(1) analysis and it is possible that the flattening seen in the CAR(1) parameter distribution for CRTS data (see Fig.~\ref{carfig}) at $\log_{10}(\tau) \sim 3$ is an artifact caused by the shorter timescale coverage relative to S82.  However, the timescale coverage of both is long enough to identify a potential characteristic timescale of $\sim150$ days in the observed frame from the Slepian wavelet variance, which we will discuss further in Sec.~\ref{timescales}.

\subsection{Color selection}

Quasars have a distinctive spectral energy distribution with a strong redshift dependency which has made color the most common selection criterion to date. With the recent availability of mid-IR all-sky photometry, selection based on {\it WISE} colors has replaced SDSS and NIR-based techniques as the most highly-rated method (although not completely reliable in selection AGN, e.g., \cite{gurkan13}). For objects with W1, W2, and W3 available, \cite{mateos12} define the following selection region: $W2 - W3 \ge 2.517$ and $ 0.315 (W2 - W3) - 0.222 < (W1 - W2) < 0.315 (W2 - W3) + 0.796$. \cite{assef12} define different criteria depending on whether the emphasis is on purity (reliability) or completeness. A purity-optimized criteria is given by:

\[ W1 - W2 > \alpha_R \exp \{\beta_R (W2 - \gamma_R)^2 \}
\]

\noindent
where $W2 < 17.11$ and  the parameters can be tuned to achieve different purity levels: $\alpha_R, \beta_R, \gamma_R = [0.662, 0.232, 13.97]$ for a purity of 90\% and $[0.530, 0.183, 13.76]$ for a purity of 75\%. For a completeness-optimized selection, a simple W1 - W2 color criterion selection suffices without any restrictions on W2: 

\[ W1 - W2 > \delta_C \]

\noindent
where $\delta_C = 0.50$ for 90\% completeness and  0.77 for 75\% completeness. This is close to the selection criterion used by \cite{stern12}:  $W1 - W2 > 0.8$ and $W2 < 15.05$. 

The Matthews correlation coefficient \citep{matthews} is a way of describing a confusion matrix with a single number: it is essentially a correlation coefficient between the observed and predicted classifications and given by:

\[ MCC = \frac{TP  \times TN - FP \times FN}{\sqrt{(TP + FP)(TP+FN)(TN+FP)(TN+FN)}} \]

\noindent
where $TP$ is the number of true positives, $TN$ is the number of true negatives, $FP$ is the number of false positives, and $FN$ is the number of true negatives. The value of MCC for the confusion matrices of the R$_{75}$, C$_{90}$ and C$_{75}$ selection criteria is approximately the same ($\sim$ 0.86) and so we take R$_{75}$ as representative.

Applying these selection criteria to our data set, we get the results shown in Table~\ref{wise}, from which we can see that {\it WISE} color selection achieves comparable results to pure variability selection. Note, however, that we have {\it WISE} colors for 98.6\% of the quasars but only 77.7\% of the stars and so the purity results must really be treated as just upper limits.

\begin{table*}
\caption{The confusion matrices derived using {\it WISE} colors to select quasars. Note, however, that {\it WISE} colors are not available for all objects in the data set and so these figures represent upper limits to the true values. MCC is the value of the Matthews correlation coefficient for the confusion matrix (see text).}
\label{wise}
\centering
\begin{tabular}{lcccccc}
\hline
 & \multicolumn{2}{c}{\bf{Mateos et al. (2012)}} & \multicolumn{2}{c}{\bf{Stern et al. (2012)}} &  \multicolumn{2}{c}{\bf{Assef et al. (2012) R$_{75}$}} \\
 & QSO & Star & QSO & Star & QSO & Star\\
\hline
QSO & 5854 & 1446 & 6107 & 1094  & 6629 & 572 \\
Star & 59 & 2272 & 6 &  2325  & 20 & 2311  \\
\hline
Completeness & \multicolumn{2}{c}{80\%} & \multicolumn{2}{c}{85\%} & \multicolumn{2}{c}{92\% }  \\
Purity & \multicolumn{2}{c}{99\%} &\multicolumn{2}{c}{99\% } &\multicolumn{2}{c}{99\% }  \\
MCC &  \multicolumn{2}{c}{0.68} &\multicolumn{2}{c}{0.76 } &\multicolumn{2}{c}{0.85} \\
$F_1$-score &  \multicolumn{2}{c}{0.89} &\multicolumn{2}{c}{0.92} &\multicolumn{2}{c}{0.96} \\
\hline
\end{tabular}
\end{table*}

\begin{figure*}
\caption{The magnitude and redshift distributions of CRTS quasars which are only detected by variability (red),  {\it WISE} color selection (R$_{75}$, black) or both (cyan).}
\label{colordep}
\includegraphics[width=3.3in]{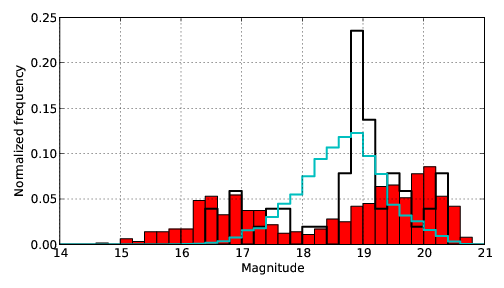}
\includegraphics[width=3.3in]{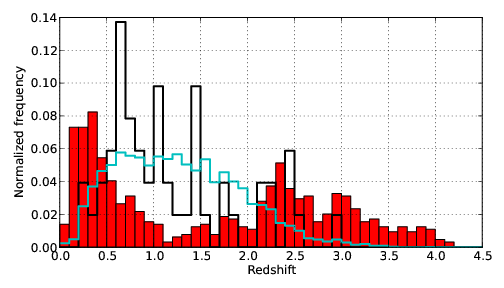}
\end{figure*}

We can determine the fraction of quasars that would be found by both a color and a variability method or just one of them. Table~\ref{colorvar} shows that about $\sim$10 -- 20\% of the quasar population, depending on exactly which color and variability criteria were used, would be missed by color selection alone and detected only by variability.Approximately 5 -- 25\% are found by color alone and not by variability respectively. Fig.~\ref{colordep} shows the magnitude and redshift distributions of the CRTS quasars which would only be detected by either color (R$_{75}$) or variability. This suggests that optical variability is more sensitive to fainter and higher redshift objects than {\it WISE} color selection but there is a (small) population of faint ($V \sim$ 19) intermediate redshift quasars that variability misses which all appear to be type 2 AGN ($r - W2 > 6$). About 30\% of this {\it WISE}-only sample are also associated with an X-ray or radio source.
With a larger sample of quasars, we will be able to better examine these different populations.

\begin{table*}
\caption{The percentages of known CRTS quasars that would be detected by just the color technique, both the color and variability technique and just the variability technique.}
\label{colorvar}
\centering
\begin{tabular}{llllllllll}
\hline
 & \multicolumn{3}{c}{{\bf Color}} & \multicolumn{3}{c}{{\bf Color + Variability}} & \multicolumn{3}{c}{{\bf Variability}} \\
 & SF & CAR(1) & SWV & SF & CAR(1) & SWV  & SF & CAR(1) & SWV \\
 \hline
Mateos et al. (2012) & 24 & 9 & 5 & 63 & 74 & 76 & 13 & 17 & 18 \\
Stern et al. (2012) & 21 & 8 & 6 & 63 & 75 & 78 & 8 & 13 & 15 \\
Assef et al. (2012) R$_{75}$ & 24 & 10 & 6 & 66 & 81 & 85 & 4 & 7 & 8 \\
\hline
\end{tabular}
\end{table*}

\subsection{Ensemble methods}

Although the individual selection techniques perform well, it should be possible to combine them in some way to construct an optimal selection method which incorporates the relative strengths of each individual one and thus obtains better predictive performance than any of them. In general, there are two types of such ensemble methods: averaging methods, such as bagging and random forests where the combined (averaged) prediction of the independent constituent methods has reduced variance; and boosting methods, in which weak component methods are combined incrementally to create a strong method with reduced bias.

We have considered four ensemble techniques in this analysis: random forests \citep{breiman}, extremely randomized trees \citep{geurts}, adaptive boosting (AdaBoost) \citep{friedman} and gradient tree boosting \citep{friedman}. For the two random tree-based methods, 30 base estimators (decision trees) were used with the Gini impurity as the splitting criterion and nodes expanded until all leaves are pure or contain less than two samples. The maximum number of features considered in looking for the best fit was also limited to the square root of the number of features presented. For the two boosting methods, 100 decision stumps and regression trees were used as the base estimator respectively with a learning rate of 1.0 in both cases. A maximum depth of three was used with the gradient tree boosting.

We have evaluated the completeness and purity achieved by each ensemble method via 10$\times$ cross-validation (see Table~\ref{ensemb_conf}) and used the individual results of the structure function ($A$ and $\gamma$), CAR(1) ($\tau$ and $\sigma^2$), and Slepian wavelet variance ($\chi^2_{qso}$ and $\chi^2_{star}$) methods as inputs. We have also considered {\it WISE} colors as additional input. 

\begin{table*}
\caption{The confusion matrices derived from the various ensemble methods using SF, CAR(1) and SWV values as input with $10 \times$ cross-validation.}
\label{ensemb_conf}
\centering
\begin{tabular}{lcccccccc}
\hline
 & \multicolumn{2}{c}{\bf{Random forest}}  & \multicolumn{2}{c}{\bf{Extremely randomized trees}} & \multicolumn{2}{c}{\bf{AdaBoost}} & \multicolumn{2}{c}{\bf{Gradient tree boosting}} \\
 & QSO & Star  & QSO & Star & QSO & Star & QSO & Star \\
\hline
QSO & 7047 & 253 & 7067 & 233 & 6975 & 325 & 6974 & 326 \\
Star & 370 & 4098 & 372 & 4118 & 454 & 4026 & 329 & 4025  \\
\hline
Completeness & \multicolumn{2}{c}{96.5\%} & \multicolumn{2}{c}{96.8\%} & \multicolumn{2}{c}{95.5\%} & \multicolumn{2}{c}{95.5\%} \\
Purity & \multicolumn{2}{c}{95.0\%} & \multicolumn{2}{c}{95.0\%} & \multicolumn{2}{c}{93.9\%} & \multicolumn{2}{c}{95.5\%}\\
$F_1$-score & \multicolumn{2}{c}{0.96} & \multicolumn{2}{c}{0.96} & \multicolumn{2}{c}{0.95} & \multicolumn{2}{c}{0.96}\\
\hline
\hline
{\bf With colors} & QSO & Star  & QSO & Star & QSO & Star & QSO & Star \\
\hline
QSO &  7148 & 53  & 7143 & 58 & 7136 & 65 & 7138 & 76 \\
Star & 75 & 2276 & 69 & 2271 & 81 & 2264 & 63 & 2266 \\
\hline
Completeness & \multicolumn{2}{c}{99.3\%} & \multicolumn{2}{c}{99.2\%} & \multicolumn{2}{c}{99.1\%} & \multicolumn{2}{c}{99.1\%} \\
Purity & \multicolumn{2}{c}{99.0\%} & \multicolumn{2}{c}{99.0\%} & \multicolumn{2}{c}{98.9\%} & \multicolumn{2}{c}{99.1\%}\\
$F_1$-score & \multicolumn{2}{c}{0.99} & \multicolumn{2}{c}{0.99} & \multicolumn{2}{c}{0.99} & \multicolumn{2}{c}{0.99}\\
\hline
\end{tabular}
\end{table*}

Although the differences are very slight, the averaging methods (random forest and extremely randomized trees) perform fractionally better than the boosting methods, both without and with colors. We have also considered all combinations of individual methods including colors and BB and find that the best results arise with SF, CAR(1), SWV and colors. In fact, we can determine the ranked contribution of each feature (method parameter) to the overall result. Although the precise rankings vary slightly for each ensemble method, the  most significant variability features are $\tau$, $\gamma$ and $\chi^2_Q$ with color, if available. 

Again it is interesting to examine the $\sim$1\% of quasars that the ensemble methods misclassify as stars.
With the caveat that we are dealing with small number statistics ($\sim$ 50 objects), most of the quasars have $W1 - W2 <  0.8$, are fainter than $V \sim 19$, have less than 150 observations in their light curves and have redshifts $< 0.7$ or $> 2$. This would suggest that they are predominantly missed because of undersampled low amplitude variability light curves. There is no association, however, with radio or X-ray sources this time.

\section{Discussion}

\subsection{Characteristic timescales}
\label{timescales}
\begin{figure*}
\caption{The rest frame Slepian wavelet variance distribution for CRTS (upper) and S82 (lower) quasars (left) and their decomposition into absolute magnitude bins (right) -- a cubic is fit to the median values within each $\tau$ bin for a given absolute magnitude range. The red data points in the left plots indicate the SWV distribution for mock light curves obeying a CAR(1) process with the same time samplings as the respective survey.}
\label{crts_restframe}
\includegraphics[width=3.45in]{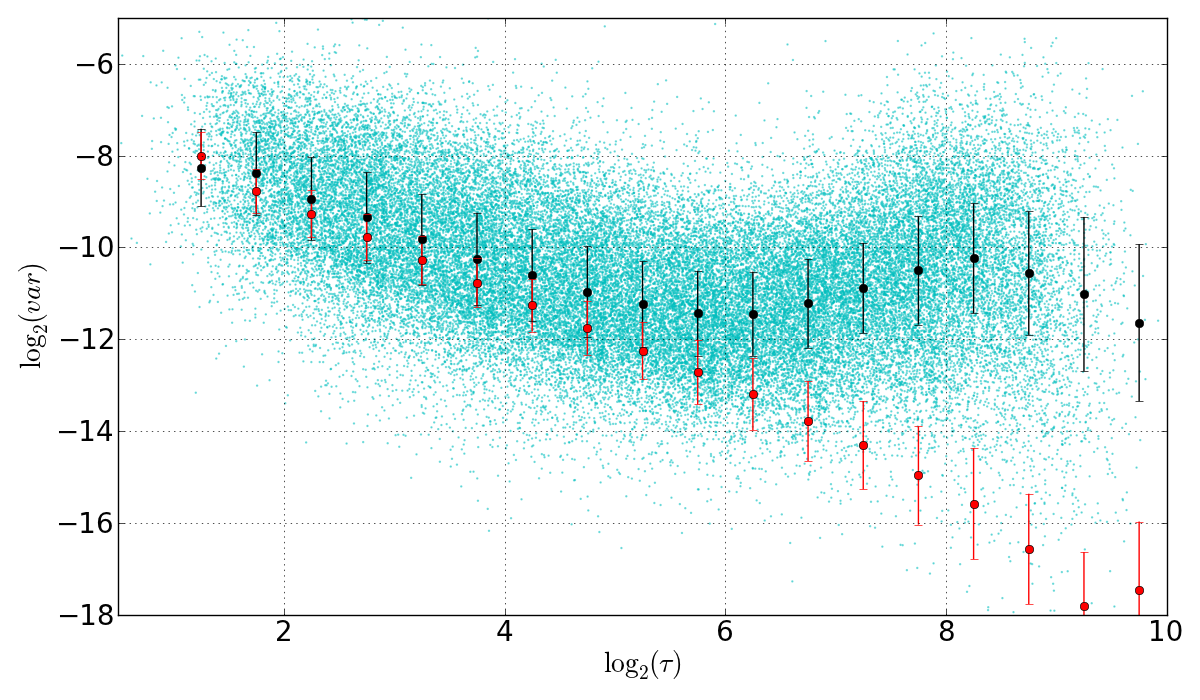}
\includegraphics[width=3.45in]{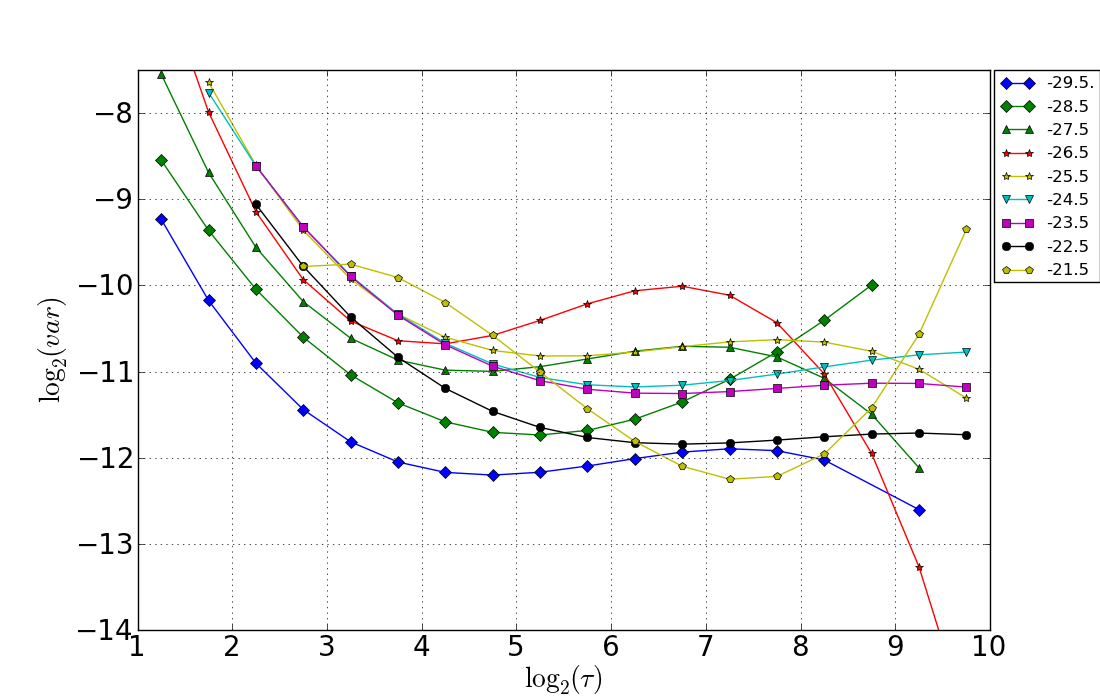}
\includegraphics[width=3.45in]{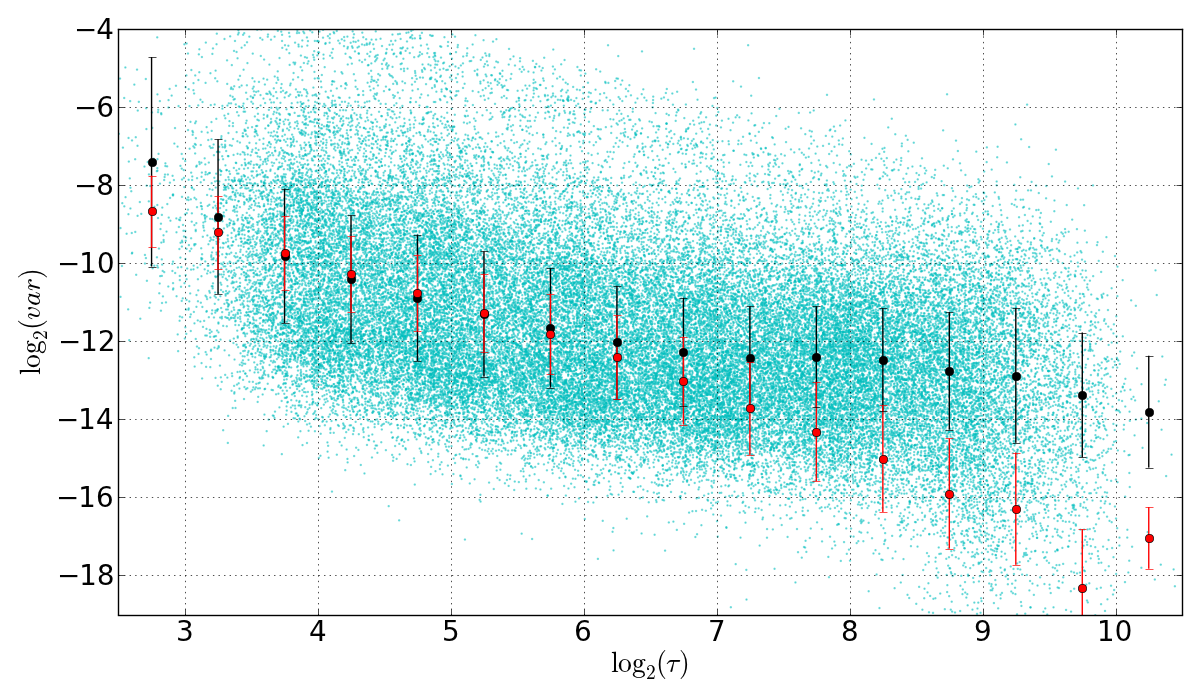}
\includegraphics[width=3.45in]{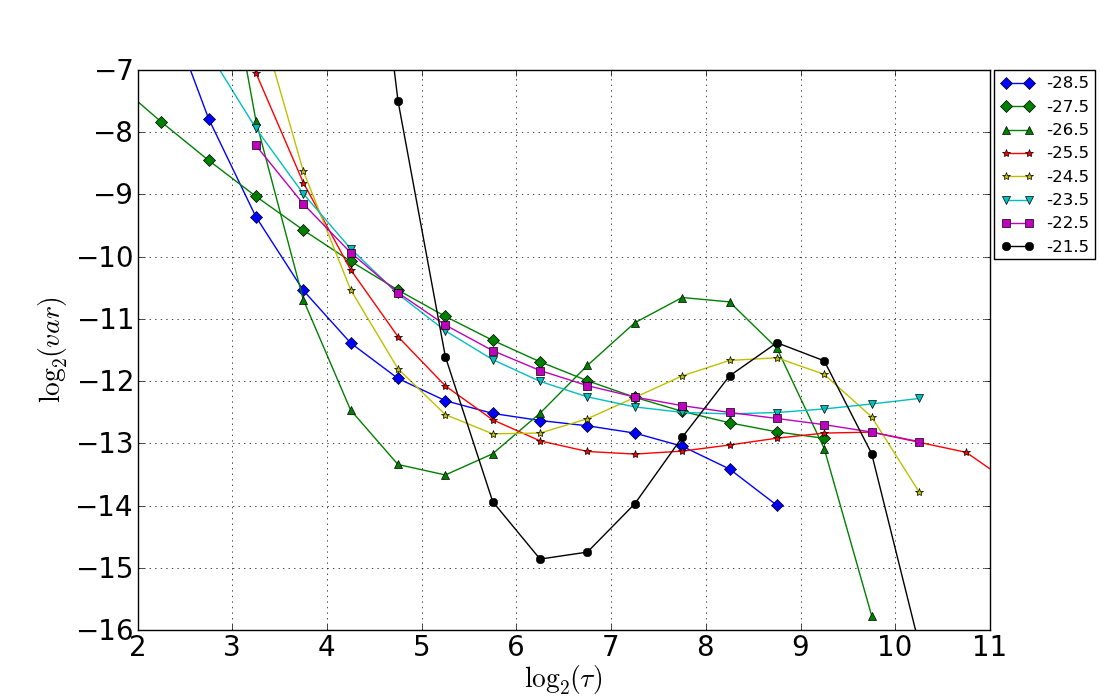}
\end{figure*}

Peaks in wavelet variance curves typically identify characteristic scales, such as periods \citep{keim10}. However, we can also consider scales at which there is a transition from one type of behaviour to another as characteristic. The slope of the log(SWV) with log(time) is indicative of different types of underlying behaviour: for example, -1 for white noise, 0 for 1/f noise, and 1 for red noise. It is clear from the CRTS quasar SWV results (see Fig.~\ref{swvcls}) that such a characteristic scale exists in the observed frame at a scale of $\log_2(\tau_o) \sim 7.25 \simeq 152$ days which is not seen in the stellar SWV distribution. We have calculated the rest frame SWV distribution for the CRTS quasars (see Fig.~\ref{crts_restframe}) and this clearly shows that the effect is intrinsic to the quasars at a rest frame scale of $\log_2(\tau_r) \sim 5.75 \simeq 54$ days. There may also be a second ``traditional'', i.e., associated with a peak, characteristic scale at $\log_2(\tau_r) \sim 8.20 \simeq 294$ days, although the sampling at this end of the SWV distribution is quite low. 

The characteristic scale at $\log_2(\tau_o) \sim 7.25$ in the observed frame is present in the Stripe 82 quasar data (see Fig.~\ref{swv82}), albeit not as obviously as with the CRTS data. However, in the rest frame SWV distribution (see Fig.~\ref{crts_restframe}), there is definitely a transition between behaviours at $\log_2(\tau_r) \sim 7.25$. The S82 SWV values are also somewhat less in scale than their CRTS counterparts which is due to the different filters used in the two samples and the known correlation of quasar variability with wavelength. 

\begin{figure}
\caption{The rest frame Slepian wavelet variance distribution for 435 MACHO/OGLE III quasars }
\label{macho}
\includegraphics[width=3.3in]{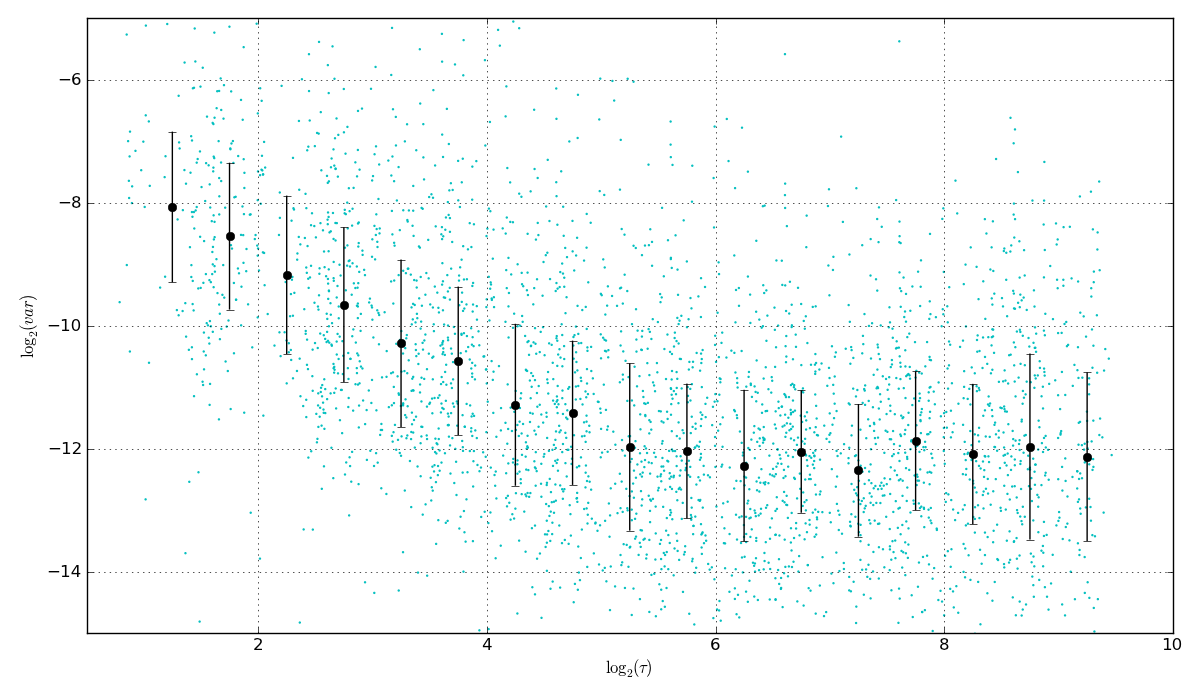}
\end{figure}

As a further check on the reality of this phenomenon, we have also computed the SWV distribution for the MACHO light curves of 435 spectroscopically confirmed OGLE III-selected quasars (see Fig.~\ref{macho}) \citep{kozlowski13}. This shows the transition between behaviours at $\log_2(\tau_r) \sim 6$ but is more consistent $\sim$1/f behaviour for $\log_2(\tau_r) > 6$. Nevertheless we see a pattern of behaviour in the SWV distribution that is consistent among quasars from three different surveys and different from that seen with stars. 

The expected behaviour for optical quasar variability is that it follows a CAR(1) model. We have generated mock light curves for both the CRTS and S82 quasar samples using the actual observation times, $t_i$, but replacing the observed magnitudes with those that would be expected under a CAR(1) model. The magnitude $X(t)$ at a given timestep $\Delta t$ from a previous value $X(t - \Delta t)$ is drawn from a Gaussian distribution with mean and variance given by \citep{kelly09}:

\[ E(X(t) | X(t - \Delta t)) = e^{-\Delta t / \tau} X(t - \Delta t) + b\tau(1 - e^{-\Delta t / \tau}) \]

\[\mathrm{Var}(X(t) | X(t - \Delta t)) = \frac{\tau \sigma^2}{2} [1 - e^{-2\Delta t / \tau}] \]

\noindent 
We add a Gaussian deviate normalized by the photometric error associated with the magnitude to be replaced at each time $t$ to incorporate measurement uncertainties into the mock light curves. For each light curve, we set $b\tau$ to its median value and use the rest frame CAR(1) best fit values determined by \cite{mac10}: $\log(\tau) = 2.305$ and $\log(SF_\infty) = -0.634$ where $SF_\infty = \tau^{1/2}\sigma$. Fig.~\ref{crts_restframe} shows the SWV distributions for the two set of mock light curves. It is clear that the transition point in both CRTS and S82 data sets indicates a rest frame scale below which quasar behaviour deviates from a CAR(1) process. 

\begin{figure}
\caption{The absolute magnitude distributions of the CRTS and S82 quasars.}
\label{absmag}
\includegraphics[width=3.3in]{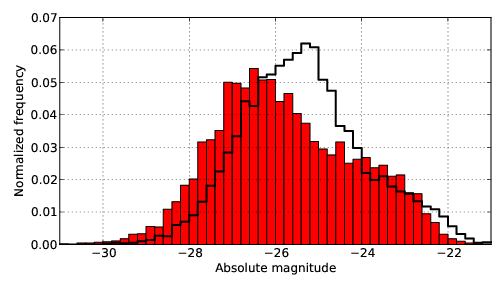}
\end{figure}

\begin{figure}
\caption{The luminosity dependence of the transition point for CRTS (cyan) and S82 (red) quasars. The dashed line indicates a weighted least-squares fit to the characteristic scales in Fig.~\ref{crts_restframe}.}
\label{transition}
\includegraphics[width=3.3in]{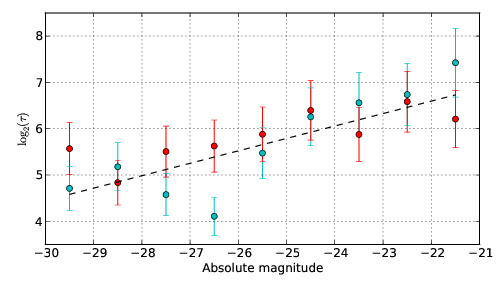}
\end{figure}

The relative position of the CAR(1) break point is different between the data sets. One distinction between them that may contribute to this is the relative fractions of quasars with different absolute magnitudes. S82 has relatively more low to mid-range ($M_{abs} > -26$) luminosity quasars than CRTS which has more luminous quasars (see Fig.~\ref{absmag}). Fig.~\ref{crts_restframe} also shows the rest frame SWV distributions in various absolute magnitude bins for CRTS and S82 quasars. There is a general anti-correlation between the strength of the variability (SWV) and absolute magnitude (in agreement with the results for $SF_{\infty}$ in \cite{mac10}) and also a luminosity dependence to the characteristic scale (transition point). The latter is shown in Fig.~\ref{transition} and suggests that the more luminous the quasar, the shorter the characteristic scale is, which can be expressed as:

\[
\log_2(\tau_r) = 0.267 * \mathrm{absolute \,\, magnitude} + 12.506
\]

\noindent
We reserve further investigations of this to a future paper.

Previous studies have identified characteristic rest frame timescales (the value of the CAR(1) $\tau$ parameter representing a damping timescale) from $\sim$200 days to 835 days \citep{kelly09, mac10, mac12}, consistent with the thermal or orbital timescales in the optically thick accretion disk of a quasar and quasar variability driven by stochastic instabilities in the accretion disk. However, \cite{zu13} detected hints of a characteristic timescale of $\sim$1--3 months in $\sim$50\% of their sample of 223 OGLE light curves (consistent with the Kepler results of \cite{mushotzky11} for four quasars) indicating a breakdown of the CAR(1) model at smaller timescales. \cite{voevodkin11} finds a best-fit solution to the structure function (SF) of Stripe 82 quasars involving a broken power law with the break at $\sim$42 days, although we note the results of \cite{emmanoulopoulos} with blazar SFs that the lengths of the data set and the shape of the underlying PS can produce spurious SF breaks. 

Studies of X-ray variability in AGN find that they are well described by a power spectrum (PS) which has a logarithmic slope, $\alpha \ge 2$, at high frequencies but flattens to a slope of $\alpha \sim 1$ below some break frequency (or corresponding characteristic timescale). Furthermore, the break frequency is seen to anti-correlate with BH mass, indicative of a correlation between the size of the X-ray emitting region and BH mass, e.g., \cite{mchardy06}. \cite{papadakis} conclude that for an AGN with a given BH mass, the accretion rate determines both the X-ray characteristic frequency and the energy spectral slope, which is consistent with models where the X-rays are produced by thermal Comptonization.  \cite{kelly11} find that there is a common anti-correlation in both optical and X-ray variability between the BH mass and the amplitude of the high-frequency PS, suggesting a shared origin for the variability, e.g., inwardly propagating fluctuations in the accretion rate. 

Assuming that quasar variability can be described by the evolution of viscous, thermal, or radiative perturbations due to a driving noise, the X-ray (high) PS break frequency corresponds to the diffusion timescale in the outer region of the accretion flow \citep{kelly11}. An additional optical (low) PS break frequency might exist at a characteristic timescale related to the time it takes a perturbation travelling at the viscous speed to cross the characteristic spatial scale of the noise field, assuming it is spatially correlated. The characteristic timescale of $\sim$54 days identified by Slepian wavelet variance is consistent with values for the low frequency break derived from fitting a mixed Ornstein-Uhlenbock process to the X-ray light curves of 10 local AGN.  Further study will help to confirm the presence of this and other characteristic timescales. 

\subsection{The control sample}

\begin{figure*}
\caption{The respective distributions for the structure function (left), CAR(1) (middle) and Slepian wavelet variance (right) applied to the CRTS control sample. The black lines in each indicate the quasar selection criteria. The red points (tracks) indicate a small set of high likelihood quasar candidates (see the text for more details).}
\label{control}
\includegraphics[width=7.0in]{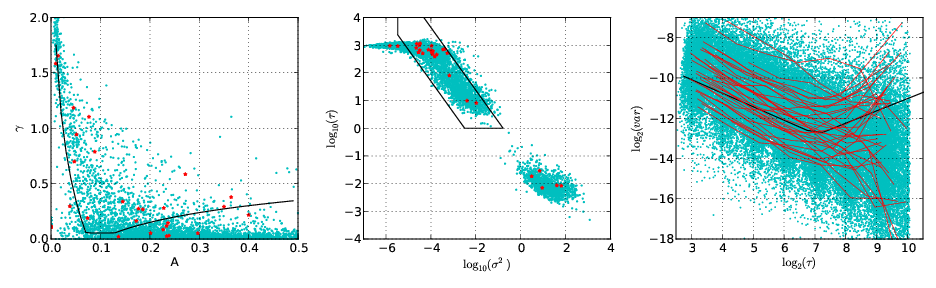}
\end{figure*}

Although we can statistically confirm the quasar selection technique that we intend to apply to the full CRTS data set, real validation only comes with spectroscopic confirmation. In Sec.~\ref{crts}, we defined a control sample of 7113 objects, which to date have not been identified as either quasars or high probability (mainly photometric) quasar candidates. This is ideal for this purpose as it is quite likely that there are as yet unidentified quasars in this data set. The distributions of the various selection algorithm parameters for this control sample are shown in Fig.~\ref{control}. Table~\ref{sfcon} gives the confusion matrices for the various selection algorithms applied to the joint set of CRTS known quasars and the control sample and shows that the QSO fraction is estimated at somewhere between 20\% and 79\% by the individual variability measures and 46\% by the ensemble technique (extremely randomized trees). Note that where the WISE color $W1- W2$ is not available, an ensemble trained on just the variability algorithm features has been used.

\begin{figure*}
\caption{Spectra of four quasar candidates observed with Keck/DEIMOS in 2013 
August.  Flux calibration was done with archival sensitivity functions 
and are only relative.  Prominent lines are labeled.  The redshift
of QC0015+2135, marked with a colon, is based on a single emission
line assumed to be Mg II, and is thus considered tentative.  All
four spectra are clearly spectroscopically confirmed as quasars, even
if the derived redshift is unclear for one.}
\label{spectra}
\includegraphics[angle=270,origin=c,width=7.0in]{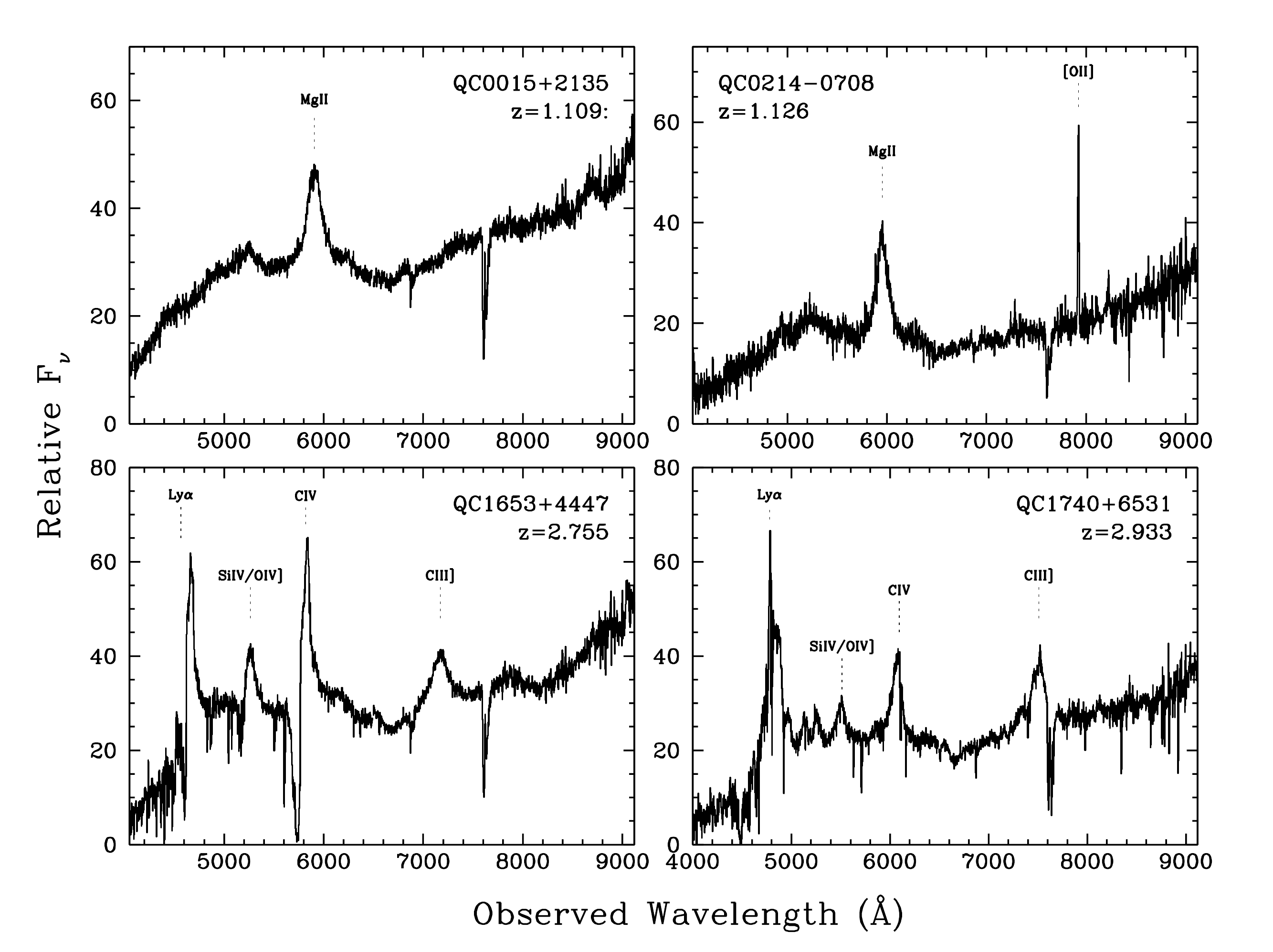}
\end{figure*}

Of the 3297 ensemble-selected quasar candidates, we have selected a representative set of 100 for spectroscopic observation to better quantify any selection biases in terms of magnitude, number of observations, {\it WISE} color, etc. Thus far, 13 of these were observed on 2013 August 30-31 with Keck + DEIMOS, one on 2013 September 3 with Palomar 200'' + DBSP, and a further six on 2013 November 8 with Palomar 200 inch + DBSP. Table~\ref{qsocand} gives details of these and their selection algorithm positions are also noted in Fig.~\ref{control}. The data were reduced using the standard DEIMOS pipeline and a representative set of spectra are shown in Fig.~\ref{spectra}. Of the 20 total candidates observed, 19 have been confirmed as quasars/AGN whilst the spectrum of the remaining one has too low a S/N to permit classification. In particular, seven have been found which previous variability techniques would not have identified. This suggests that our ensemble classification technique should produce a highly complete quasar catalog.

\begin{table*}
\caption{The confusion matrices derived from the various algorithms to identify quasars in the CRTS control sample. Note that completeness, purity and $F_1$-scores for BB11 do not include objects with ambiguous classifications.}
\label{sfcon}
\centering
\begin{tabular}{lcccccccccc}
\hline
 & \multicolumn{2}{c}{\bf{SF}}  & \multicolumn{2}{c}{\bf{CAR(1)}} & \multicolumn{2}{c}{\bf{BB11}} & \multicolumn{2}{c}{\bf{SWV}} & \multicolumn{2}{c}{\bf{Ensemb}}\\
 & QSO & Control & QSO & Control & QSO & Control & QSO & Control & QSO & Control \\
\hline
QSO & 5154 & 2143 & 6433 & 629 & 4915 & 1474 & 6854 & 446 & 7294 & 3 \\
Control & 1394 & 5718 & 4003 & 2747 & 1570 & 3820 & 5585 & 1528  & 3297 & 3770\\
\hline
Completeness & \multicolumn{2}{c}{71\%} & \multicolumn{2}{c}{91\%} & \multicolumn{2}{c}{77\%} & 
\multicolumn{2}{c}{94\%} & \multicolumn{2}{c}{100\%} \\
Purity & \multicolumn{2}{c}{79\%} & \multicolumn{2}{c}{62\%} & \multicolumn{2}{c}{76\%} & \multicolumn{2}{c}{55\%} & \multicolumn{2}{c}{69\%}\\
$F_1$-score &  \multicolumn{2}{c}{0.74} & \multicolumn{2}{c}{0.74} & \multicolumn{2}{c}{0.76} & \multicolumn{2}{c}{0.69} & \multicolumn{2}{c}{0.82}\\
\hline
\end{tabular} 
\end{table*}

\begin{table*}
\caption{Details of the 20 quasar candidates that were observed. In the predicted columns: `Q' is a quasar, `-' is not a quasar and `?' indicates ambiguous for BB11 alone. A colon on the redshift indicates that it is based on a single emission line assumed to be Mg II and is thus considered tentative. Note that the S/N for QC0938+1957 was too low to permit a classification.}
\label{qsocand}
\centering
\begin{tabular}{llllllllllll}
\hline
 & & & & & & \multicolumn{5}{c}{Predicted} & \\
Id  & RA & Dec & $V$ & $W1 - W2$ & W2 & SF & CAR(1) & BB11 & SWV & Ensemble &z \\
\hline
$QC0015+2135^a$ &  00 15 17.65 & $+21$ 35 06.40 & 18.62 & 1.26 & 13.69& Q & Q & Q & Q & Q & 1.109: \\ 
$QC0017+1104^a$ &  00 17 14.63 & $+11$ 04 19.10 & 19.57 &  1.45 & 14.03&  Q  & - & - & Q & Q & 1.186: \\ 
$QC0105+2755^b$ &  01 05 16.30 & $+27$ 55 23.30 &19.67 & 0.88 & 14.04& - & Q & - & Q & Q & 0.536 \\ 
$QC0150+0328^a$ &  01 50 58.88 & $+03$ 28 22.70 & 19.36 & 1.11  & 14.22& Q & Q & - & Q &Q  & 1.821 \\ 
$QC0213-0708^a$ &  02 13 46.80 & $-07$ 08 22.20 & 20.52 & 1.09 & 14.99& Q & Q & - & Q & Q & 1.126 \\ 
$QC0226+0429^a$ & 02 26 26.38 & $+04$ 29 41.20 & 18.01 & 1.05 & 12.91& Q & - & ? & Q & Q & 1.471 \\ 
$QC0326-0534^c$ &  03 26 51.00 & $-05$ 34 53.47 & 18.65 & 0.85 & 14.27 & - & - & - & Q & Q &  1.618 \\ 
$QC0352-1731^c$ &  03 32 56.92 & $-17$ 31 47.39 & 19.63 & 1.08 & 14.17 & - & - & - & Q & Q & 0.817 \\ 
$QC0745+2438^c$ & 07 45 40.44 & $+24$ 38 01.28 & 18.86 &0.84 & 14.84& - & - & - & Q & Q & 0.242 \\ 
$QC0938+1957^c$ &  09 38 58.91& $+19$ 57 34.31 & 19.26 &0.95 &14.12 & - & - & - & Q & Q & - \\ 
$QC1024+1842^c$ &  10 24 29.55 & $+18$ 42 50.69 & 18.27 & 1.16& 13.74& - & - & - & Q & Q & 0.187 \\ 
$QC1653+4447^a$ & 16 53 27.81 & $+44$ 47 02.70 & 18.79 & 0.95 &14.72 & Q & Q & Q & Q & Q & 2.755 \\ 
$QC1706+1318^a$ &  17 06 02.51 & $+13$ 18 46.10 & 17.79 & 0.62 & 15.76& - & Q & Q & Q & Q& 0.095: \\ 
$QC1740+6531^a$ &  17 40 53.24 & $+65$ 31 11.50 & 19.19 & 1.10 & 14.99& Q & - & ? & Q & Q & 2.933 \\ 
$QC1741+4432^a$ &  17 41 09.91 & $+44$ 32 27.00 & 18.96 & 1.21 & 14.96& - & - & - & Q & Q & 2.328 \\ 
$QC2136-0221^a$ &  21 36 00.37 & $-02$ 21 10.30 & 19.73 & 1.20 & 14.88& Q & Q & Q & Q & Q & 1.735 \\  
$QC2313+3502^c$ &  23 13 18.22 & $+35$ 02 40.27 & 18.76 &0.97 & 14.47& - & - & - & Q & Q &  0.240\\ 
$QC2324-0052^a$ & 23 24 28.36 & $-00$ 52 44.30 & 19.48 & 1.00 &13.76 & - & - & - & Q & Q & 0.506 \\ 
$QC2339+2923^a$ & 23 39 41.69 & $+29$ 23 59.20 & 18.39 & 1.43 &13.31 &  Q  & Q & Q & Q & Q &  1.630 \\ 
$QC2346+2523^a$ &  23 46 40.47 & $+25$ 23 31.10 & 18.42 & 1.12& 12.95 & Q & Q & - & Q & Q & 0.783 \\ 

\hline
\end{tabular}
a: observed on Aug 30/31 2013 with Keck + DEIMOS; 
b : observed on Sep 3 2013 with Palomar 200" + DBSP;
c: observed on Nov 8 2013 with Palomar 200'' + DBSP.
\end{table*}

\section{Conclusions}

The production of a large quasar sample selected on the basis of quasar variability is viable now. Previous investigations have focused on SDSS Stripe 82 or OGLE data, whilst looking forward to Pan-STARRS and LSST for larger well-sampled data sets. However, the CRTS data set is perfect for quasar studies with an estimated coverage of almost 200,000 known quasars, 500,000 photometric quasar candidates and a conservatively estimated further 500,000 new quasars amongst its 500 million light curves.

In this paper, we have compared the efficacy of common quasar variability-based selection techniques on CRTS data and also introduced a new technique based on Slepian wavelet variance. We find that this is as good as (if not better than) the other techniques but makes fewer assumptions. It is also able to identify characteristic timescales within a time series, either those contributing significantly more to the overall variance of the object or those associated with a change of behaviour. 

We find strong evidence from a combined total of 18028 CRTS, SDSS and MACHO quasars for a transition timescale of $\sim 54$ days in the quasar rest frame from the behaviour predicted by a CAR(1) model. This confirms the short timescale deviation hinted at in OGLE data by \cite{zu13} and {\it Kepler} data by \cite{mushotzky11}. If multiple mechanisms are responsible for quasar variability or it is the result of a single mechanism but with a combination of different timescales and amplitudes, one should expect to see deviations from a CAR(1) model. Such a break might also be expected in models where variability is driven by a spatially correlated noise field.

Subsequent analyses of larger CRTS quasar samples using the fuller DR2 data will allow a more thorough investigation of this timescale and its dependencies on physical parameters such as luminosity, black hole mass and Eddington limit. We also intend to extend the Slepian wavelet variance technique to support finer time resolution and explore whether quasars are better described by more sophisticated continuous time stochastic processes than just CAR(1). Finally, we note that combining CRTS data with other existing quasar time series should also prove a profitable avenue for quasar studies.

\section*{Acknowledgments}

We thank the anonymous reviewer for their comments and Debashis Mondal, Donald Percival, Kaspar Schmidt, Nathaniel Butler, Brandon Kelly and Nathalie Palanque-Delabrouille for useful discussions, data and code. We also thank the staff of the Keck and Palomar Observatories for their assistance with observations.

This work was supported in part by the NSF grants AST-0909182, IIS-1118041 and AST-1313422, by the W. M. Keck Institute for Space Studies, and by the U.S. Virtual Astronomical Observatory, itself supported by the NSF grant AST-0834235. 

This work made use of the Million Quasars Catalogue.

This publication makes use of data products from the {\it Wide-field Infrared Survey Explorer}, which is a joint project of the University of California, Los Angeles, and the Jet Propulsion Laboratory/California Institute of Technology, funded by the National Aeronautics and Space Administration.

The authors wish to recognize and acknowledge the very significant cultural role and reverence that the summit of Mauna Kea has always had within the indigenous Hawaiian community.  We are most fortunate to have the opportunity to conduct observations from this mountain.

Funding for SDSS-III has been provided by the Alfred P. Sloan Foundation, the Participating Institutions, the National Science Foundation, and the U.S. Department of Energy Office of Science. The SDSS-III web site is http://www.sdss3.org/.

SDSS-III is managed by the Astrophysical Research Consortium for the Participating Institutions of the SDSS-III Collaboration including the University of Arizona, the Brazilian Participation Group, Brookhaven National Laboratory, Carnegie Mellon University, University of Florida, the French Participation Group, the German Participation Group, Harvard University, the Instituto de Astrofisica de Canarias, the Michigan State/Notre Dame/JINA Participation Group, Johns Hopkins University, Lawrence Berkeley National Laboratory, Max Planck Institute for Astrophysics, Max Planck Institute for Extraterrestrial Physics, New Mexico State University, New York University, Ohio State University, Pennsylvania State University, University of Portsmouth, Princeton University, the Spanish Participation Group, University of Tokyo, University of Utah, Vanderbilt University, University of Virginia, University of Washington, and Yale University.

\appendix

\section[]{Slepian wavelet variance}
\label{appa}

Following \citep{mondal11}, we consider an observed time series $X(t_0), X(t_1), \ldots, X(t_{N-1})$ with data taken at irregular time points $t_0, t_1, \ldots,  t_{N+1}$. The sampling intervals are $\Delta_1 = t_1 - t_0, \Delta_2 = t_2 - t_1, \ldots, \Delta_n = t_n - t_{n-1}$, which gives an average sampling interval of:

\[
\bar{\Delta} = \frac{1}{N-1} (\Delta_1 + \Delta_2 + \ldots + \Delta_{N-1})  = \frac{t_{N-1} - t_0}{N-1} 
\]

Now for a fixed positive integer $j$, we define the passband of frequencies

\[ 
A_j = [-2^{-j}/ \bar{\Delta}, -2^{-j-1}/\bar{\Delta}]  \,\, \cup \,\, [2^{-j-1} / \bar{\Delta}, 2^{-j} / \bar{\Delta}] 
\]

\noindent
and dyadic scales $\tau_j = \bar{\Delta} 2^{j-1}$ for $j = 1, 2, \ldots, J$ with $N+1 = 2^J + N_J, 0 < N_J \leq 2^J -1$.  For each $k$, let $\{\psi_{k,m}\}^{M-1}_{m=0}$ be the coefficients of a linear filter that approximates a bandpass filter with passband $A_j$ and is adaptive to time points $t_k, t_{k+1}, \ldots, t_{k+M-1}$. Considering the Fourier transform:

\[
\Psi_k(f) = \sum_{m=0}^{M-1} \psi_{k,m} e^{-i2\pi ft_{k_m}}
\]

\noindent
we desire $\{\psi_{k,m}\}$ such that: 

\begin{enumerate}
\item filter coefficients sum to zero: $\sum_m \psi_{k,m} = 0$
\item the sum of the squares of the coefficients are normalized: $\sum_m \psi_{k,m}^2 = 1 / 2^{-j} \bar{\Delta}$  
\item the squared gain function $|\Psi_k(f)|^2$ is as concentrated as possible within $A_j$, i.e., the linear filter maximizes the energy contained in the passband:

\[
\max_{\psi} \lambda(k, M, j) = \frac{\int_{A_j} |\Psi_k(f)|^2 df}{\int^{1/2}_{1/2} |\Psi_k(f)|^2 df} = \frac{\psi^T_kQ_{k,j}\psi_k}{\psi^T_k\psi_k} \]

\noindent
where the $(s,t)$th element of the $M \times M$ matrix $Q_{k,j}$ is:

\[ 
Q_{k,j} (m,m') = \int_{A_j} e^{-i 2 \pi f(t_{k+m} - t_{k+m'})} df \]
\[
 =  \frac{\sin \left(\frac{2\pi(t_{k+m} - t_{k+m'})}{2^j\bar{\Delta}} \right) - \sin \left(\frac{2\pi(t_{k+m} - t_{k+m'})}{2^{j+1}\bar{\Delta}} \right)}{\pi(t_{k+m} - t_{k+m'})} \] 

\end{enumerate}
 
We want to find the maximum eigenvector $Q_{k,j}\psi_k = \lambda(k, M, j)\psi_k$ subject to $1^T\psi_k = 0$. 
Setting $M=c2^j$ where $2c$ is an integer independent of $j$ and assuming that the sampling intervals arise from a stationary process, for large $j$, this becomes a continuous eigenvalue problem:

\[ \int_{-1}^{1} \mu \beta_c(f - f') \psi(f') df' = \lambda \psi_k(f) \]

\noindent
where $f = 2m / M -1$, $f' = 2m' / M -1$ so that $-1 \le f, f' < 1$ and $\mu = E \bar{\Delta}$. The adaptive filters can then be obtained from the continuous Slepian wavelet $\psi(f)$ via:

\[ 
\psi \left( 2 \frac{t_{m+k} - t_k}{t_{M+k} - t_k} -1 \right), \, m = 0, 1, \ldots, M-1, \,\, k = 0, 1, \ldots \]

The Slepian wavelet coefficients indexed by scale $\tau_j$ and shift $k$ are defined by:

\[ U_{j,k} = \sum_{u=0}^{M_j -1} \psi_{j,k,u} X(t_{k+u}), k = 0, 1, \ldots, N - M_j \]

\noindent
and an approximate estimate of the wavelet variance at scale $\tau_j$ is:

\[ \hat{\nu}^2 (\tau_j) = \frac{1}{N - M_j + 1} \sum_{k=0}^{N-M_j} U^2_{j,k} \]

\noindent
Under appropriate regularity conditions, $\hat{\nu}^2 (\tau_j) $ is asymptotically a Gaussian random variable.

\label{lastpage}

\end{document}